%Implications of N=1 Superconformal Symmetry for Chiral Fields
%Plain TeX file using harvmac

\input harvmac
%\draft
\input amssym.def
\input amssym
\baselineskip 14pt
\magnification\magstep1
\parskip 6pt
%For smaller font footnotes
\newdimen\itemindent \itemindent=32pt
\def\textindent#1{\parindent=\itemindent\let\par=\resetpar%
\indent\llap{#1\enspace}\ignorespaces}

\let\oldpar=\par
\def\resetpar{\oldpar\parindent=20pt\let\par=\oldpar}

\font\ninerm=cmr9 \font\ninesy=cmsy9
\font\eightrm=cmr8 \font\sixrm=cmr6
\font\eighti=cmmi8 \font\sixi=cmmi6
\font\eightsy=cmsy8 \font\sixsy=cmsy6
\font\eightbf=cmbx8 \font\sixbf=cmbx6
\font\eightit=cmti8
\def\eightpoint{\def\rm{\fam0\eightrm}
  \textfont0=\eightrm \scriptfont0=\sixrm \scriptscriptfont0=\fiverm
  \textfont1=\eighti  \scriptfont1=\sixi  \scriptscriptfont1=\fivei
  \textfont2=\eightsy \scriptfont2=\sixsy \scriptscriptfont2=\fivesy
  \textfont3=\tenex   \scriptfont3=\tenex \scriptscriptfont3=\tenex
  \textfont\itfam=\eightit  \def\it{\fam\itfam\eightit}%
  \textfont\bffam=\eightbf  \scriptfont\bffam=\sixbf
  \scriptscriptfont\bffam=\fivebf  \def\bf{\fam\bffam\eightbf}%
  \normalbaselineskip=9pt
  \setbox\strutbox=\hbox{\vrule height7pt depth2pt width0pt}%
  \let\big=\eightbig  \normalbaselines\rm}
\catcode`@=11 %
\def\eightbig#1{{\hbox{$\textfont0=\ninerm\textfont2=\ninesy
  \left#1\vbox to6.5pt{}\right.\n@@space$}}}
\def\vfootnote#1{\insert\footins\bgroup\eightpoint
  \interlinepenalty=\interfootnotelinepenalty
  \splittopskip=\ht\strutbox %
  \splitmaxdepth=\dp\strutbox %
  \leftskip=0pt \rightskip=0pt \spaceskip=0pt \xspaceskip=0pt
  \textindent{#1}\footstrut\futurelet\next\fo@t}
\catcode`@=12 %
\def \d{{\rm d}}
\def \de{\delta}
\def \si{\sigma}

\def \pr{\partial}
\def \d{{\rm d}}
\def \tr{{\rm tr }}
\def \ta{{\tilde a}}

\def \hx{{\hat {\rm x}}}

\def \bs{{\bar s}}

\def \bI{{\bar I}}
\def \bJ{{\bar J}}

\def \l{\langle}
\def \r{\rangle}
\def \ep{\epsilon}

\def \half{{\textstyle {1 \over 2}}}
\def \thir{{\textstyle {1 \over 3}}}
\def \quar{{\textstyle {1 \over 4}}}
\def \ts{\textstyle}

\def \C{{\cal C}}

\def \F{{\cal F}}

\def \I{{\cal I}}
\def \L{{\cal L}}
\def \N{{\cal N}}

\def \tb{{\tilde {\rm b}}}
\def \tx{{\tilde {\rm x}}}

\def \tsi{{\tilde \sigma}}
\def \bth{{\bar\theta}}
\def \bTh{{\bar \Theta}}
\def \bkap{{\bar{\kappa}}}
\def \bta{{\bar \eta}}
\def \bep{{\bar \epsilon}}
\def \bphi{{\bar \phi}}
\def \bom{{\bar{\omega}}}
\def \hom{{\hat{\omega}}}
\def \bX{{\bar{X}}}
\def \teta{{\tilde \eta}}
\def \tth{{\tilde \theta}}
\def \dal{{\dot \alpha}}
\def \dbe{{\dot \beta}}

\def \a{{\rm a}}
\def \b{{\rm b}}

\def \x{{\rm x}}

\def \Y{{\rm Y}}
\def \hY{{\hat{\rm Y}}}
\def \1{{\bar 1}}
\def \2{{\bar 2}}
\def \3{{\bar 3}}

\font \bigbf=cmbx10 scaled \magstep1

%References
\lref\hughtwo{J. Erdmenger and H. Osborn, Nucl. Phys. {B483} (1997)
431, hep-th/9605009.}
\lref\hughone{H. Osborn and A. Petkou,
    Ann. Phys. (N.Y.) {231} (1994) 311, hep-th/9307010.}
\lref\Wess{J. Wess and J. Bagger, {\it Supersymmetry and Supergravity}
(Princeton University Press, Princeton, 1983).}
\lref\Buch{I.L. Buchbinder and S.M. Kuzenko, {\it Ideas and Methods of
Supersymmetry and Supergravity} (IOP Publishing Ltd., Bristol, 1995).}
\lref\Buch{I.L. Buchbinder and S.M. Kuzenko, {\it Ideas and Methods of
Supersymmetry and Supergravity} (IOP Publishing Ltd., Bristol, 1995).}
\lref\Symanzik{K. Symanzik, Lett. al Nuovo Cimento 3 (1972) 734.}
\lref\Pick{A. Pickering and P. West, Nucl. Phys. B569 (2000) 303,
hep-th/9904076.}
\lref\HO{H. Osborn, Ann. Phys. (N.Y.) 272 (1999) 243, hep-th/9808041.}
\lref\Park{J-H. Park, Nucl. Phys. B539 (1999) 599, hep-th/9807176\semi
J-H. Park, Nucl. Phys. B559 (1999) 455, hep-th/9903230.}
\lref\Erd{{\it Higher Transcendental Functions, vol. I}, ed. A. Erd\'elyi, 
(McGraw-Hill Book Co., New York, 1953), p. 222-242.}
\lref\Ferr{S. Ferrara, A.F. Grillo, R. Gatto and G. Parisi, Nuovo Cimento 19A
(1974) 667.}
\lref\CFT{P. Di Francesco, P. Mathieu and D. S\'en\'echal, {\it Conformal Field
Theory}, (Springer, New York, 1996).}
\lref\Ext{H. Exton, J. Physics A 28 (1995) 631.}
\lref\Ape{P. Appell and J. Kamp\'e de F\'eriet, {\it Fonctions hyperg\'eometriques and
hypersph\'eriques: polynomes d'Hermite} (Gauthier-Villars, Paris, 1926).}
\lref\Kuz{S.M. Kuzenko and S. Theisen, Class. and Quant. Gravity 17 (2000) 665,
hep-th/9907107.}
\lref\Depp{M. D'Eramo, G. Parisi and L. Peliti, Lett. al Nuovo Cimento 2 (1971) 878.}
\lref\Lang{K. Lang and W. R\"uhl, Nucl. Phys. {B377} (1992) 371.}
\lref\Dot{Vl.S. Dotsenko, Nucl. Phys. B235 [FS11] (1984) 54.}
\lref\Uss{N.I. Ussyukina and A.I. Davydychev, Phys. Lett. B298 (1993) 363; 
B305 (1993) 136.}
\lref\Dav{A.I. Davydychev and J.B. Tausk, Nucl. Phys. B397 (1993) 133.}
\lref\Pet{A.C. Petkou, Ann. Phys. (N.Y.) 249 (1996) 180, hep-th/9410093.}
\lref\Fone{S. Ferrara, A.F. Grillo, R. Gatto and G. Parisi, Nucl. Phys. 
B49 (1972) 77.}
\lref\West{E. D'Hoker, D.Z. Freedman, S.D. Mathur, A. Matusis,
L. Rastelli, hep-th/9908160.\semi
B. Eden, P.S. Howe, C. Schubert, E. Sokatchev and P.C. West,
Phys. Lett. B472 (2000) 323, hep-th/9910150\semi
B. Eden, P.S. Howe, E. Sokatchev and P.C. West, hep-th/0004102\semi
M. Bianchi and S. Kovacs, Phys. Lett. B468 (1999) 102; hep-th/9910016\semi
J. Erdmenger and M. P\'erez-Victoria, Phys. Rev. D62 (2000) 045008,
hep-th/9912250\semi
B. Eden, C. Schubert and E. Sokatchev, Phys. Lett. B482 (2000) 309, 
hep-th/0003096\semi
E. D'Hoker, J. Erdmenger, D.Z. Freedman and M. P\'erez-Victoria, 
hep-th/0003218.}
\lref\WCon{B.P. Conlong and P.C. West, J. Phys. A 26 (1993) 3325.}

{\nopagenumbers
\rightline{DAMTP/00-59}
\rightline{hep-th/0006098}
\vskip 1.5truecm
\centerline {\bigbf Implications of $\N=1$ Superconformal Symmetry for
Chiral Fields}
\vskip  6pt
%\centerline
\vskip 2.0 true cm
\centerline {F.A. Dolan and H. Osborn${}^\dagger$}

\vskip 12pt
\centerline {\ Department of Applied Mathematics and Theoretical Physics,}
\centerline {Silver Street, Cambridge, CB3 9EW, England}
\vskip 1.5 true cm

{\eightpoint
\parindent 1.5cm{

{\narrower\smallskip\parindent 0pt
The requirements of $\N=1$ superconformal invariance for the correlation
functions of chiral superfields are analysed. Complete expressions are found
for the three point function for the general spin case and for the four
point function for scalar superfields for $\sum q_i=3$ where  $q_i$
is the scale dimension for the $i$'th superfield and is related to
the $U(1)$ $R$-charge. In the latter case the relevant 
Ward identities reduce to eight differential equations for four functions of
$u,v$ which are invariants when the superconformal symmetry is reduced
to the usual conformal group. The differential equations have a general
solution given by four linearly independent expressions involving
a two variable  generalisation of the hypergeometric function. By
considering the behaviour under permutations, or crossing symmetry, the 
chiral four point function is shown to be determined up to a single overall 
constant. The results are in accord with the supersymmetric operator
product expansion.

\narrower}}

\vfill
\line{${}^\dagger$ 
address for correspondence: Trinity College, Cambridge, CB2 1TQ, England\hfill}
\line{\hskip0.2cm emails:
{{\tt fad20@damtp.cam.ac.uk} and \tt ho@damtp.cam.ac.uk}\hfill}
}
%\vskip0.5cm

%PACS: 11.10-z; 11.25.Hf; 11.10Kk; 04.62+v
\eject}
\pageno=1

\newsec{Introduction}

It is now clear that there exist in four, and also five and six,  dimensions
a plethora of non trivial quantum field theories, without any coupling to gravity,
which possess fixed points realising conformal symmetry. The evidence for such
conformally invariant theories is strongest in the case where the field theory
is supersymmetric and the conformal group is extended to the superconformal
group. In four dimensions this may be identified as $SU(2,2|\N)$ where the
cases of $\N=1,2,4$ are relevant for renormalisable field theories, although
for $\N=4$ the group contains an ideal so the group may be reduced to the
projective group $PSU(2,2|4)$.
Besides  the usual 15 parameter conformal group the superconformal group also
contains an $R$-symmetry, $U(1)$, $U(2)$, and $SU(4)$ for $\N=1,2$ and $4$. The 
case of $\N=4$ gauge theories has been known for a long time since these have 
vanishing $\beta$-functions and recently the strong coupling limit of such
theories has been explored through the ADS/CFT correspondence. Nevertheless
$\N=1$ theories may possess, for appropriate matter content, a conformal window
with superconformal invariant fixed points.

Such theories are naturally described in terms of superfields defined over
a superspace involving Grassman variables. In general the action of conformal
transformations on the coordinates is non linear and the construction of covariant
correlation functions is not as straightforward as realising the consequencies
of symmetry under the usual Poincar\'e
group transformations  which act linearly on the coordinates. 
A formalism which allows the construction of two and three point, and
in principle $n>3$ as well, correlation functions was described in \HO\ for the
$\N=1$ superconformal group in four dimensions, extending a similar discussion
for the ordinary conformal group in \refs{\hughone,\hughtwo}.
This has been extended to six dimensions and also to $\N>1$ in four dimensions by 
\Park\ and this approach has been
used for an analysis of particular two and three point functions
for $\N=2$ in \Kuz. In four dimensions the superconformal group is compatible with
acting on superfields restricted to the chiral projections, $z_+=(x_+, \theta)$ 
where $\theta^\alpha$ is a chiral spinor, or $z_-=(x_-,\bth)$, where $\bth^\dal$
is an anti-chiral spinor. For two points $z_1 = (x_1, \theta_1 , \bth_1) $ and
$z_2 = (x_2, \theta_2 , \bth_2) $ then the constructions described in \HO\ are
expressed in terms of $\tx_{\1 2}= x_{\1 2}{}^{\!\! a}\tsi_a$  which is 
constructed from $z_{1-}$ and $z_{2+}$ and transforms
homogeneously under superconformal transformations with a local scale 
transformation and rotation at $z_1$ and also at $z_2$, in particular
\eqn\twop{
x_{\1 2}{}^{\! 2} \longrightarrow  {x_{\1 2}{}^{\! 2}\over {\bar \Omega}(z_{1-})
\Omega(z_{2+})} \, ,
}
which is a direct generalisation of the transformation of $(x_1-x_2)^2$ under
the usual conformal group. For three points in superspace $z_1,z_2,z_3$ it is
also possible to generalise the treatment for forming conformally covariant
expressions \refs{\hughone,\hughtwo} by  introducing the variables
$X_1, \Theta_1$ and their conjugates $\bX_1 , \bTh_1$, which are
related in the same fashion as $x_+, \theta$ and $x_- , \bth$, that transform
homogeneously under local rotations and scale transformations at $z_1$. 
In particular $X_1{}^{\! 2} = x_{\2 3}{}^{\! 2} /(x_{\2 1}{}^{\! 2} \,
x_{\1 3}{}^{\! 2})$ and its transformation follows from \twop. These
variables play an essential role in the general construction of three point
functions described in \HO.

However $X_1$ depends on $z_1, z_{2-}, z_{3+}$ while $\Theta_1$ is formed from
$z_1, z_{2-}, z_{3-}$ and correspondingly for their conjugates.
In consequence the construction of correlation
functions for chiral fields alone, which depend only on the $z_+$'s, requires
cancellations of unwanted chiral components in the formalism of \HO. This 
was achieved quite easily for the three point function for chiral scalar
superfields $\l \phi_1(z_{1+}) \phi_2(z_{2+}) \phi_3(z_{3+}) \r$
but efforts to generalise such a treatment to the four point function became 
very involved and were not carried through to a conclusion. In a different
approach Pickering and West \Pick\ obtained identities expressing 
superconformal invariance directly for three and four point functions which 
from the start depend solely on the $z_+$'s.\foot{The three point function was 
considered earlier by Conlong and West \WCon.} The three point case was again
rather simple but the invariance for the four point function led to a system
of eight linear first order partial differential equations for four functions of
two variables $u,v$ which are the two independent cross ratios
formed from $(x_{i+} - x_{j+})^2$. Without supersymmetry $u,v$ are conformal
invariants and the general conformally covariant four point function involves
an arbitrary function of $u,v$.

In this paper we follow a similar path to finding the conditions following
from $\N=1$ superconformal invariance
for chiral superfield correlation functions. For such superfields the scale
dimension $q$ is trivially related to the $R$-charge $3q$ and they may belong
only to $(j,0)$ spinor representations of the four dimensional Lorentz group.
For a subgroup of $Sl(4|1)$, the
complexification of $SU(2,2|1)$, which  contains the conformal group, the
constraints are simply realised where by introducing, for three points 
$z_{1+}, z_{2+}, z_{3+}$, spinor variables $\Lambda$ which transform
homogeneously. Extending this to the full superconformal group involves further
constraints. For the three point function we must require $\sum q_i = 3$
and for the four point
function, with the same condition on the $q_i$'s, we obtain eight differential 
equations similar to those found by Pickering and West. 
These are here solved in terms of a certain
two variable generalisation of the hypergeometric function $F_4$.
The superconformal symmetry identities for the correlation function
for four scalar chiral superfields have in general four linearly 
independent solutions.  A careful analysis of the behaviour under permutations
of the superfields or imposing crossing symmetry, which requires taking into 
account non trivial identities for
the $F_4$ functions under analytic continuation, leads to a unique expression 
up to an overall constant.

In this paper in the next section we describe how the the superconformal
group acts on the chiral coordinates $z_+$ and obtain the essential 
transformation formulae used to construct superconformal correlation functions
later. In section three we derive the superconformal identities for the
three and four point functions for chiral fields. Results for the three
point function are obtained for arbirary spin. For the four point function
we restrict to scalar fields and the eight equations for four functions of
$u,v$ are obtained. 
These equations have four linearly independent solutions which are expressed
in terms of the functions $F_4$ mentioned above.  The conditions necessary 
for Bose symmetry or crossing symmetry are analysed in section 4.
By requiring independence under the path of analytic continuation, and using
the transformation formulae for the $F_4$ function, a unique
result, up to a single overall constant, is obtained.
In section 5 an alternative expression in terms of a function $G$, defined
in terms of two $F_4$ functions, is obtained. This allows the short distance
limits of the four point function to become manifest. In section 6 we
endeavour to count the constraints for higher point functions of chiral
scalar superfields. At least for six and higher point functions the
differential equations arising from superconformal invariance do not fully
determine the functions arising in the general expansion. In section 7
we consider the application of the operator product expansion for two
chiral superfields to the four point function. The contribution of 
a chiral scalar superfield in the operator product expansion to the
four point function is obtained in a form which satisfies all the 
superconformal Ward identities but which does not satisfy the crossing
symmetry relations by itself.  In section 8  we show how
an integral over superspace defines a superconformal $N$-point function
for any $N$ and this is shown to be in accord with the results obtained
in section 3 for $N=3,4$.
In appendix A we discuss an alternative derivation of the essential
equations for the four point function based on using superconformal
transformations to fix two points. In appendix B we describe some 
results relevant in section 8 for the evaluation of conformal integrals
while in appendix C some results which allow for the simplification
of the two variable function $G$ used here are described. For the
cases of relevance here it is shown how they may be reduced to sums
of products of ordinary hypergeometric functions. In appendix D we
recapitulate some results for the operator product expansion which
are applied to the chiral superfield four point function in appendix E.

\newsec{Superconformal Transformations on Chiral Superspace}

We use the standard identification of 4-vectors with $2\times 2$ 
matrices\foot{The notation is identical with \HO\ and is essentially that of
of Wess and Bagger \Wess. Thus $ \theta^\alpha, \,
\bth^\dal$ are regarded as row, column vectors and we let
$\tth_\alpha = \ep_{\alpha\beta}\theta^\beta, \, {\tilde \bth}{}_\dal
= \ep_{\smash {\dal \dbe}} \bth{}^\dbe$ form associated column, row vectors,
$\theta^2 = \theta \tth, \, \bth^2 = {\tilde \bth}\bth$. The
basis of $2\times 2$-matrices is given by  the hermitian $\si$-matrices
$\si_a, \, \tsi_a, \, \si_{(a} \tsi_{b)} = - \eta_{ab}1$, and for a 4-vector
$x^a$ then  $\x_{\alpha\dal} =
x^a (\si_a)_{\alpha\dal}, \ \tx^{\dal\alpha} = x^a (\tsi_a)^{\dal\alpha}
= \ep^{\alpha\beta}\ep^{\smash {\dal \dbe}} \x_{\smash {\beta \dbe}}$,
with inverse $x^a = - {1\over 2}{\rm tr}(\si^a \tx)$.}
so that the action of infinitesimal superconformal transformations on $z_+$
is given by \refs{\Buch,\HO}
\eqn\sol{ \!\!\!\!\!\!\eqalign{
\de \tx_+ ={}& \ta  + \bom \tx_+ - \tx_+ \omega  + (\kappa+\bkap) \tx_+ 
+ \tx_+ \b \tx_+ + 4i \, \bep \theta - 4\, \tx_+ \eta \theta  \, , \cr
\de \theta ={}& \ep - \theta \omega + \kappa \, \theta + 
\theta \b \tx_+  -i  \bta \,\tx_+ + 2\, \teta \, \theta^2 \, ,  \cr
\omega_\beta{}^\alpha = {}& -\quar \omega^{ab}
(\si_a \tsi_b)_\beta{}^\alpha \, ,  \qquad
\bom{}^\dal{}_{\smash\dbe} =
-\quar \omega^{ab} (\tsi_a \si_b)^\dal{}_{\smash\dbe}\, , \cr}
}
where $a^a$ corresponds to  a translation, $\omega^{ab}=-\omega^{ba}$ an 
infinitesimal 
rotation, $b^a$ a special conformal transformation, $\kappa+\bkap$ a rescaling
and $(\kappa-\bkap)/i$ a $U(1)_R$ phase while $\ep^\alpha, \bep^\dal$ are 
supertranslations with $\eta_\alpha,\bta_{\smash \dal}$ their superconformal 
extensions. These parameters may be written as elements of a supermatrix
\eqn\MM{
M = \pmatrix{\omega - {\ts{1\over3}}(\kappa + 2\bkap)1 &
- i \b & 2\eta \cr -i {\tilde \a} & \bom + {\ts{1\over3}}(2\kappa + \bkap ) 1& 
2\bep \cr 2 \ep & 2 \bta & {\ts{2\over3}} ( \kappa - \bkap ) } \, ,
}
so that
\eqn\lie{
\de_2 \de_1 - \de_1 \de_2 = \de_3 \qquad \Rightarrow \qquad
[ M_1 , M_2 ] = M_3 \, .
}
Since ${\rm str}\, M =0$, $M$ is a generator of $Sl(4|1)$ which is reduced
to $SU(2,2|1)$ by imposing the appropriate reality condition.

For two points $z_{i+},z_{j+}$ we may define
\eqn\defxl{
\tx_{ij} = \tx_{i+} - \tx_{j+} \, , \qquad \ell_{ij} = (\theta_i - \theta_j)
\tx_{ij}{}^{\!\! -1} \, , \quad \tx_{ij}{}^{\!\! -1} = \x_{ji}/x_{ij}{}^{\! 2} \, ,
}
which, from \sol, transform as
\eqn\txl{\eqalign{
\de \tx_{ij} = {}& \big ( \bom + \bkap + \tx_{i+}\b + 4i\,  \bep_i \ell_{ij}
\big ) \tx_{ij} 
+ \tx_{ij} \big ( - \omega + \kappa + \b \tx_{j+} - 4 \, \eta \theta_j \big )
\, , \cr
\de \ell_{ij} = {}& - \ell_{ij} \big ( \bom + \bkap + \tx_{i+} \b - 4\,  \theta_i
\eta \big ) - i \bta + \theta_i \b + 2i \, \ell_{ij}{}^{\! 2}\, 
{\tilde \bep}_i \, , \cr}
}
where we also define
\eqn\defhbe{
\bep(x_{+}) = \bep + i\,  \x_{+} \bta \, , \qquad \bep_i = \bep(x_{i+}) \, .
}
Furthermore if
\eqn\defrl{
r_{ij} = (x_{i+} - x_{j+})^2 = - \det(\tx_{ij}) \, , 
}
defining also the spinor $\Lambda_{i(jk)}$ and 4-vector $Y_{i(jk)}$ by
\eqn\defLY{
\Lambda_{i(jk)} = \ell_{ij} - \ell_{ik} \, , \qquad
\Y_{i(jk)} = - \tx_{ij}{}^{\!\! -1} + \tx_{ik}{}^{\!\! -1} = \tx_{ji}{}^{\!\! -1}
\tx_{jk}\, \tx_{ik}{}^{\!\! -1}  = - \tx_{ki}{}^{\!\! -1}\tx_{kj}\, \tx_{ij}{}^{\!\! -1}
\, ,
}
we then have
\eqn\tr{
\de r_{ij} = 2 \big ( \kappa + \bkap - b{\cdot x}_i - b{\cdot x}_j + 2 \theta_j
\eta - 2i \, \ell_{ij} \bep_i\big ) r_{ij} \, ,
}
and
\eqn\tL{\eqalign{
\de \Lambda_{i(jk)} = {}& - \Lambda_{i(jk)}\big ( \bom + \bkap + \tx_{i+} \b 
+ 4\,  \theta_i \eta  -2i \, ({\tilde \ell}_{ij} +
{\tilde \ell}_{ik} ) {\tilde \bep}_i \big ) \, , \cr
\de \Y_{i(jk)} = {}& \big ( \omega  - \kappa - \b \tx_{i+} + 4\eta \theta_i \big )
\Y_{i(jk)} \cr 
&{} - \Y_{i(jk)} \big ( \bom + \bkap + \tx_{i+} \b + 4i \,
\bep_i \ell_{ij} - 4i \, \tx_{ij} \tx_{kj}{}^{\!\! -1} \bep_k \, \Lambda_{i(jk)}
\big ) \, . \cr }
}
If we set $\bep, \eta = 0$  then it is evident from \txl\ that
$\tx_{ij}$ varies homogeneously with local scale transformations and rotations
at $z_{i+}$ and $z_{j+}$.  Similarly the variations of $\Lambda_{i(jk)}$ 
and $Y_{i(jk)}$ given by $\tL$
then corresponds to a local scale transformation and rotation at $z_{i+}$. 
The group $G_0 \subset Sl(4|1)$ generated by matrices $M$ as in \MM\ with
$\bep, \eta = 0$ contains the usual bosonic conformal group and it is
straightforward to construct covariantly transforming correlation functions
and also invariants under $G_0$. Thus for four points $z_{1+},z_{2+},z_{3+},
z_{4+}$ we may define the $G_0$ invariant cross ratios
\eqn\defuv{
u= {r_{12} \, r_{34} \over r_{13} \, r_{24}} \, , \qquad \qquad
v= {r_{14} \, r_{23} \over r_{13} \, r_{24}} \, .
}
For later use we also define unit 4-vectors by
\eqn\defYx{
\hx_{ji} = \x_{ji}\, r_{ij}{}^{\!\! -{1\over 2}} = 
\tx_{ij}{}^{\!\! -1}r_{ij}{}^{\! {1\over 2}} \, , \qquad
\hY_{i(jk)} = \Y_{i(jk)} \bigg({r_{jk}\over r_{ij} \, r_{ik}} 
\bigg )^{\! -{1\over 2}} \, ,
}
so that, from \txl\ and \tL, we have
\eqn\txY{\eqalign{
\de \hx_{ji} = {}& \hom_j \hx_{ji} -  \hx_{ji} \big ( \bom + \half ( \tx_i \b - \tb \x_i)
+ 4i ( \bep_i \ell_{ij} + \half \ell_{ij}\bep_i \, 1) \big ) \, , \cr
\de \hY_{i(jk)}  = {}& \hom_i \hY_{i(jk)} -  \hY_{i(jk)}  \big ( \bom +
\half ( \tx_i \b - \tb \x_i) + 4i ( \bep_i \ell_{ij} + \half \ell_{ij}\bep_i \, 1) \cr
& \qquad\qquad\qquad\quad - 4i\, (\tx_{ij} \tx_{kj}{}^{\!\! -1} \bep_k  \Lambda_{i(jk)} 
+ \half \Lambda_{i(jk)} \tx_{ij} \tx_{kj}{}^{\!\! -1} \bep_k \, 1) \big ) \, , \cr}
}
where $\hom_\alpha{}^\beta(z_+)$, $\hom_\alpha{}^\alpha=0$, is given by
\eqn\defw{
\hom(z_+) = \omega - \half (  \b \tx_{+} - \x_+ \tb) + 4 (\eta \theta +
\half \theta \eta \, 1 ) \, , \qquad \hom_i = \hom(z_{i+}) \, .
}
In obtaining \txY\ it is useful to note, with the definitions in \defhbe,
\eqn\tbep{
\bep_i = \tx_{ij}  \tx_{kj}{}^{\!\! -1} \bep_k +
\tx_{ik} \tx_{jk}{}^{\!\! -1} \bep_j \, .
}

To achieve symmetry under the full superconformal group requires cancellation
of all terms involving $\bep_i$. This requires further constraints. To obtain
the necessary conditions we first obtain from \tL,
\eqn\tLL{
\de \Lambda_{i(jk)}{}^{\! 2} = -2\big ( \bkap - b{\cdot x}_{i+} + 4\,  
\theta_i \eta
+ i (\ell_{ij} + \ell_{ik}) \bep_i \big ) \Lambda_{i(jk)}{}^{\! 2} \, ,
}
so that, using \tr, we have
\eqn\tLr{
\de \, {\Lambda_{i(jk)}{}^{\! 2} \over r_{ij}} = - 2(2\si_i + \si_j ) 
{\Lambda_{i(jk)}{}^{\! 2}  \over r_{ij}} \, ,
}
with
\eqn\defsi{
\si(z_+) = {\ts{1\over 3}}(\kappa+2\bkap)- b{\cdot x_{+}} + 2\, \theta \eta  
\, , \qquad \si_i = \si(z_{i+}) \, .
}
In \tLr\ the terms involving $\bep_i$ have therefore cancelled. For subsequent
use we also note that, from \tr\ and \defsi,
\eqn\trr{
\de \, {r_{ij} \over r_{ik}} = 2 \big ( \si_j - \si_k - 2i \, \Lambda_{i(jk)}
\bep_i\big ) { r_{ij} \over r_{ik}} \, .
}

{}From the definition in \defrl\ we have
\eqn\LL{
\Lambda_{i(jk)} = \Lambda_{i(jl)} + \Lambda_{i(lk)} \, , \qquad
\Lambda_{i(jk)} = - \Lambda_{i(kj)} \, ,
}
as well as
\eqn\LLx{
\Lambda_{i(jk)} = \Lambda_{j(ik)} \tx_{jk} \tx_{ik}{}^{\!\! -1} \, .
}
This leads to
\eqn\SL{
{\Lambda_{i(jk)}{}^{\! 2} \over r_{jk}} = {\Lambda_{j(ki)}{}^{\! 2} \over r_{ik}}
= {\Lambda_{k(ij)}{}^{\! 2} \over r_{ij}} \, ,
}
so that $\Lambda_{i(jk)}{}^{\! 2} / r_{jk}$  is completely symmetric. This
is manifest in the transformation properties, obtained from  \tLr\ and \trr,
\eqn\tLL{
\de \, {\Lambda_{i(jk)}{}^{\! 2} \over r_{jk}} = - 2 (\si_i + \si_j + \si_k )
{\Lambda_{i(jk)}{}^{\! 2} \over r_{jk}} \, .
}

These results allow the construction of superconformal covariant expressions
for chiral superfield correlation functions. For a quasi-primary
spin $j$ chiral field belonging to the $(j,0)$ representation
$\phi_I = \phi_{\alpha_1 \dots \alpha_{2j}}$, totally symmetric in
$\alpha_1 \dots \alpha_{2j}$ so that $I$ takes $2j+1$ values, we have,
with $\hom$ and $\si$ defined in \defw\ and \defsi,
\eqn\superj{ \eqalign{
\de \phi_I(z_+) = {}& - \L_{z_+} \phi_I(z_+) - 2q \, \si(z_+) \phi_I(z_+)
+ (\hom(z_+) {\cdot s} )_I{}^J  \phi_J(z_+)  \, , \cr
\L_{z_+} = {}& {\de z_+}{\cdot {\pr \over \pr z_+}} \, ,\qquad
(\hom {\cdot s} )_I{}^J \phi_J  =2j \, \hom_{(\alpha_1}{}^{\!\beta}
\phi_{\alpha_2 \dots \alpha_{2j})\beta} \, , \cr}
}
where $\hom {\cdot s}  = \hom_\beta{}^\alpha s_\alpha{}^\beta$ with 
$s_\alpha{}^\beta$, $[ s_\alpha{}^\beta , s_\gamma{}^\delta ] =  
\de_\alpha{}^{\!\delta} s_\gamma{}^\beta - 
\de_{\gamma}{}^{\!\beta} s_\alpha{}^\delta $, generators for spin $j$.
The representations are specified by $(j,q)$ and for these to be unitary 
with positive energy
either $j=q=0$, which is the trivial singlet case, or $q\ge j+1$, with $q$
determining both the $R$-charge and scale dimension.

\newsec{Chiral Three and Four Point Functions}

The superconformal Ward identities for the $n$-point chiral correlation 
function of chiral superfields belonging to representations $(j_i,q_i)$, 
$i=1,\dots n$, requires
\eqn\scf{ \eqalign{
\sum_i \big ( & \L_{i} + 2q_i \, \si_i \big ) \big \l \phi_{I_1}(z_{1+}) \dots 
\phi_{I_n}(z_{n+}) \big \r \cr
&  - \sum_i (\hom_i{\cdot s_i})_{I_i}{}^{\! J} \big \l \phi_{I_1}(z_{1+}) \dots 
\phi_{J}(z_{i+}) \dots \phi_{I_n}(z_{n+}) \big \r  = 0 \, , \cr}
}
with $s_i$ the generators for spin $j_i$.
We solve these identities explicitly for general spins when $n=3$  and
for scalar chiral superfields when $n=4$ so long as
\eqn\sumq{
\sum_i q_i = 3 \, .
}
\bigskip

For the three point function we may write
\eqn\tphi{
 \big \l \phi_{I_1}(z_{1+}) \phi_{I_2}(z_{2+}) \phi_{I_3}(z_{3+}) \big \r =
{\Lambda_{1(23)}{}^{\! 2} \over r_{23}} \, 
\bigg ( {r_{12} \over r_{13}} \bigg )^{\!  1- q_2} 
\bigg ( {r_{12} \over r_{23}} \bigg )^{\! 1- q_1} \!
F_{I_1 I_2 I_3} (x_{1+},x_{2+},x_{3+} ) \, .
}
For scalar fields $F_{I_1 I_2 I_3} \to C_{123}$, a constant, 
and it is easy to see from \tLL\ and
\trr, since $\Lambda_{1(23)}{}^{\! 3}=0$, that this has the correct transformation
properties in accord with \scf\ and also from \SL, with \sumq, it is completely
symmetric under simultaneous permutations of $z_{1+},z_{2+},z_{3+}$ and 
$q_1,q_2,q_3$. To construct $F_{I_1 I_2 I_3}$ for non zero spins we introduce
\eqn\defI{
\I^{(j)}({\hat x})_{I \bI} = \hx_{(\alpha_1|\dal_1} \dots 
\hx_{\alpha_{2j}) \dal_{2j}} \, ,
}
which is a bi-spinor, transforming as $(j,0) \times (0,j)$. The appropriate
solution is then
\eqn\solF{
F_{I_1 I_2 I_3}(x_{1+},x_{2+},x_{3+} ) = \I^{(j_1)}({\hat Y}_{1(23)})_{I_1 \bI_1} 
\I^{(j_2)}({\hat x}_{21})_{I_2 \bI_2} \I^{(j_3)}({\hat x}_{31})_{I_3 \bI_3} \,
t^{ \bI_1 \bI_2 \bI_3}_{\, j_1j_2j_3} \, , 
}
where $\hat{x}_{21}, \hat{x}_{31}$ and ${\hat Y}_{1(23)}$ are given by \txY\
and $t^{ \bI_1 \bI_2 \bI_3}_{\, j_1j_2j_3}$ is an invariant tensor satisfying, 
for any rotation $r$,
\eqn\invt{
D^{(j_1)}(r)^{\bI_1}{}_{\! \bJ_1} D^{(j_2)}(r)^{\bI_2}{}_{\! \bJ_2} 
D^{(j_3)}(r)^{\bI_3}{}_{\! \bJ_3}\, t^{ \bJ_1 \bJ_2 \bJ_3}_{\, j_1j_2j_3} = 
t^{ \bI_1 \bI_2 \bI_3}_{\,j_1j_2j_3} \, .
}
Such tensors are given by the usual Clebsch-Gordan coefficients if $|j_2-j_3|
\le j_1 \le j_2+j_3$. Infinitesimally \invt\ requires
$(\bom {\cdot \bs}_1)^{\bI_1}{}_{\! \bJ}\, t^{ \bJ\, \bI_2 \bI_3}_{\,j_1j_2j_3}
+ (\bom {\cdot \bs}_2)^{\bI_2}{}_{\! \bJ}\, t^{ \bI_1 \bJ \bI_3}_{\,j_1j_2j_3}
+ (\bom {\cdot \bs}_3)^{\bI_3}{}_{\! \bJ}\, t^{ \bI_1 \bI_2 \bJ}_{\,j_1j_2j_3}
= 0$, for any $\bom^\dal{}_\dbe, \,\bom^\dal{}_\dal=0$,   and using 
this with \txY, and since terms 
$\propto \Lambda_{1(23)}$ in \txY\ vanish, it is straightforward to see that
this ensures that \solF\ has
the correct spinorial transformation properties according to \scf. Although
the construction in \solF\ is apparently asymmetric we may re-express
the result in the equivalent form expected by permutation symmetry.
Using \invt\ first with
$D^{(j)}(r) = \I^{(j)}(\hat x_{32})^{-1} \I^{(j)}(\hat x_{31})$ then we have
$\I^{(j_1)}({\hat Y}_{1(23)}) = \I^{(j_1)}( \hat x_{12} )
D^{(j_1)}(r)$, since $\hY_{1(23)} = \hx_{12} \hx_{32}{}^{\!\! -1} \hx_{31}$, 
and similarly $\I^{(j_2)}( \hat Y_{2(13)}) = \I^{(j_2)}({\hat x}_{21})
D^{(j_2)}(r)^{-1}$, or secondly with
$D^{(j)}(r) = \I^{(j)}(\hat x_{23})^{-1} \I^{(j)}(\hat x_{21})$ when we have
$\I^{(j_1)}({\hat Y}_{1(23)}) = (-1)^{2j_1} \I^{(j_1)}( \hat x_{13} )
D^{(j_1)}(r)$, allows the expression \solF\ also to be written as
\eqn\solFt{\eqalign{ 
F_{I_1 I_2 I_3}(x_{1+},x_{2+},x_{3+} ) = {}&\I^{(j_1)}(\hat x_{12})_{I_1 \bI_1}
\I^{(j_2)}({\hat Y}_{2(13)})_{I_2 \bI_2} \I^{(j_3)}({\hat x}_{32})_{I_3 \bI_3}\,
t^{ \bI_1 \bI_2 \bI_3}_{\, j_1j_2j_3} \, , \cr
= {}&\I^{(j_1)}(\hat x_{13})_{I_1 \bI_1}
\I^{(j_2)}(\hat x_{23})_{I_2 \bI_2} \I^{(j_3)}({\hat Y}_{3(12)})_{I_3 \bI_3}\,
(-1)^{2j_1} t^{ \bI_1 \bI_2 \bI_3}_{\, j_1j_2j_3} \, . \cr}
}

Fermi/Bose symmetry following from $\phi_1(z_{1+})\phi_2(z_{2+}) = P_{12}
\phi_2(z_{2+})\phi_1(z_{1+})$, where $P_{12}=-1$ for both $2j_1,2j_2$ odd, 
and with similar definitions of $P_{23}, P_{13}$, requires then
\eqn\FB{
P_{23} t^{ \bI_1 \bI_3 \bI_2}_{\, j_1j_3j_2} = (-1)^{2j_1} 
t^{ \bI_1 \bI_2 \bI_3}_{\, j_1j_2j_3} \, , \quad
P_{12} t^{ \bI_2 \bI_1 \bI_3}_{\, j_2j_1j_3} = 
t^{ \bI_1 \bI_2 \bI_3}_{\, j_1j_2j_3} \, , \quad
P_{13}P_{23} t^{ \bI_3 \bI_1 \bI_2}_{\, j_3j_1j_2} = (-1)^{2j_1}
t^{ \bI_1 \bI_2 \bI_3}_{\, j_1j_2j_3} \, .
}
If $(j_1,q_1)=(j_2,q_2)$ this provides constraints on possible values of
$j_3$, and similarly for any other identical pair.

The discussion of the four point function is more involved and so is restricted
here to chiral scalar superfields $\phi_i$, $i=1,2,3,4$. As shown in the 
section 6, and in accord with the analysis of Pickering and West \Pick,
we may restrict to an expansion in a  nilpotent basis formed by 
$\Lambda_{i(jk)}{}^{\! 2}$, $1\le i < j < k \le 4$, with coefficients
involving functions of the cross ratios $u,v$ defined in \defuv.
Thus we write,
\eqn\fourp{\eqalign{
\big \l \phi_1(z_{1+}) \phi_2 & (z_{2+}) \phi_3(z_{3+}) \phi_4(z_{4+}) \big \r \cr
= {}& {\Lambda_{1(23)}{}^{\! 2} \over r_{23}}  
\bigg ( {r_{12} \over r_{13}} \bigg )^{\! q_3 -1 } \!
\bigg ( {r_{23} \over r_{13}} \bigg )^{\! q_1 + q_4 - 1} \!
\bigg ( {r_{12} \over r_{24}} \bigg )^{\! q_4 } \! f_{123}(u,v) \cr
{}& + {\Lambda_{1(24)}{}^{\! 2} \over r_{24}}    
\bigg ( {r_{14} \over r_{24}} \bigg )^{\! q_2 -1 } \!
\bigg ( {r_{12} \over r_{24}} \bigg )^{\! q_3 + q_4 - 1} \!
\bigg ( {r_{14} \over r_{13}} \bigg )^{\! q_3 } \! f_{124}(u,v) \cr
{}& + {\Lambda_{1(34)}{}^{\! 2} \over r_{34}}    
\bigg ( {r_{34} \over r_{13}} \bigg )^{\! q_1 -1 } \!
\bigg ( {r_{14} \over r_{13}} \bigg )^{\! q_2 + q_3 - 1} \!
\bigg ( {r_{34} \over r_{24}} \bigg )^{\! q_2 } \! f_{134}(u,v) \cr
{}& + {\Lambda_{2(34)}{}^{\! 2} \over r_{34}}    
\bigg ( {r_{23} \over r_{24}} \bigg )^{\! q_4 -1 } \!
\bigg ( {r_{34} \over r_{24}} \bigg )^{\! q_1 + q_2 - 1} \!
\bigg ( {r_{23} \over r_{13}} \bigg )^{\! q_1 } \! f_{234}(u,v) \, .\cr}
}
It is easy to see that this has the correct form to satisfy the superconformal
Ward identities up to terms proportional to $\bep_i$ which are generated
by the variation of $u,v$ and also by the variation of the last factor of 
$r_{ij}/r_{ik}$ in
each of the four terms appearing in the expression for the four point
function given by \fourp. Such terms must cancel. To achieve a linearly
independent basis we restrict all contributions to only $\bep_1, \bep_2$, by
using \tbep, and also to involve just $\Lambda_{1(23)}{}^{\! 2}\Lambda_{1(24)}$
and $\Lambda_{1(24)}{}^{\! 2}\Lambda_{1(23)}$. Using \trr\ and \LLx\ we may find
\eqn\varv{ \eqalign{
\de v = {}& - 4iv ( \Lambda_{1(43)} \bep_1 + \Lambda_{2(34)} \bep_2 ) \cr
= {}& - 4iv \big ( \Lambda_{1(23)} ( \bep_1 - \tx_{13} \tx_{23}{}^{\!\! -1} 
\bep_2 ) - \Lambda_{1(24)} ( \bep_1 - \tx_{14} \tx_{24}{}^{\!\! -1}\bep_2 ) 
\big ) \, , \cr}
}
and also 
\eqnn\varu
$$\eqalignno{
\de u = {}& - 4iu ( \Lambda_{1(23)} \bep_1 + \Lambda_{4(32)} \bep_4 ) \cr
= {}& - 4i \big ( \Lambda_{1(23)} - \Lambda_{1(24)} \big ) \big ( 
u \bep_1 - v \, \tx_{13} \tx_{23}{}^{\!\! -1} \bep_2 + 
\tx_{14} \tx_{24}{}^{\!\! -1} \bep_2
- u \, \tx_{13} \tx_{43}{}^{\!\! -1}\tx_{42} \tx_{12}{}^{\!\! -1}\bep_1 \big ) \cr
{}& - 4iu \, \Lambda_{1(24)} \tx_{14} \tx_{24}{}^{\!\! -1}\bep_2 \, . & \varu
\cr}
$$
It is easy to see that
\eqn\LLL{
\Lambda_{1(34)}{}^{\! 2} \Lambda_{1(23)} = \Lambda_{1(34)}{}^{\! 2}
\Lambda_{1(24)} = \Lambda_{1(23)}{}^{\! 2}\Lambda_{1(24)} +
\Lambda_{1(24)}{}^{\! 2}\Lambda_{1(23)} \, ,
}
and, from \varv\ and \varu, we have
\eqn\varlv{\eqalign{
\Lambda_{2(34)}{}^{\! 2} \de v = {}& 4i v \, {r_{13}\over r_{23}}
\Lambda_{1(23)}{}^{\! 2}\Lambda_{1(24)} \big ( \bep_1 - 
\tx_{14} \tx_{24}{}^{\!\! -1}\tx_{23} \tx_{13}{}^{\!\! -1}\bep_1 \big )\cr
{}& - 4i v \, {r_{14}\over r_{24}}
\Lambda_{1(24)}{}^{\! 2}\Lambda_{1(23)} \big ( \bep_1 -
\tx_{13} \tx_{23}{}^{\!\! -1}\tx_{24} \tx_{14}{}^{\!\! -1}\bep_1 \big )\, , \cr}
}
and
\eqn\varlu{
\Lambda_{2(34)}{}^{\! 2} \de u = - 4iu \, {r_{13}\over r_{23}}
\Lambda_{1(23)}{}^{\! 2}\Lambda_{1(24)} 
\tx_{14} \tx_{24}{}^{\!\! -1}\tx_{23} \tx_{13}{}^{\!\! -1}\bep_1 
- 4iu \, {r_{14}\over r_{24}} \Lambda_{1(24)}{}^{\! 2}\Lambda_{1(23)} \bep_1 \, .
}
To complete the calculation we need the contributions from the variations
of the relevant $r_{ij}/r_{ik}$ factors in the four terms appearing in \fourp. 
These are given respectively by
\eqn\varr{\eqalign{
-4iq_4 \Lambda_{1(23)}{}^{\! 2} \Lambda_{2(14)}\bep_2 = {}& 
-4iq_4 \Lambda_{1(23)}{}^{\! 2} \Lambda_{1(24)} \tx_{14} \tx_{24}{}^{\!\! -1}
\bep_2 \, ,\cr
-4iq_3 \Lambda_{1(24)}{}^{\! 2} \Lambda_{1(43)}\bep_1 = {}& 
-4iq_3 \Lambda_{1(24)}{}^{\! 2} \Lambda_{1(23)}\bep_1 \, , \cr
-4iq_2 \Lambda_{1(34)}{}^{\! 2} \Lambda_{4(32)}\bep_4  = {}&
4iq_2 \big ( \Lambda_{1(23)}{}^{\! 2}\Lambda_{1(24)} +
\Lambda_{1(24)}{}^{\! 2}\Lambda_{1(23)} \big ) \big ( \bep_1 -
\tx_{14} \tx_{24}{}^{\!\! -1} \bep_2 \big ) \, , \cr
-4iq_1 \Lambda_{2(34)}{}^{\! 2} \Lambda_{3(21)}\bep_3  = {}& 
-4iq_1 \, {r_{13}\over r_{23}}\Lambda_{1(23)}{}^{\! 2}\Lambda_{1(24)}
\big ( \tx_{14} \tx_{24}{}^{\!\! -1}\tx_{23} \tx_{13}{}^{\!\! -1}\bep_1 -
\tx_{14} \tx_{24}{}^{\!\! -1} \bep_2 \big ) \cr
{}& - 4iq_1 \, {r_{14}\over r_{24}} \Lambda_{1(24)}{}^{\! 2}\Lambda_{1(23)}
\big ( \bep_1 - \tx_{13} \tx_{23}{}^{\!\! -1} \bep_2 \big ) \, .  \cr}
}
Using in addition the identities
\eqn\xxxx{\eqalign{
\tx_{14} \tx_{24}{}^{\!\! -1}\tx_{23} \tx_{13}{}^{\!\! -1} = {}& u\,
\tx_{13} \tx_{43}{}^{\!\! -1}\tx_{42} \tx_{12}{}^{\!\! -1} + v-u \, , \cr
v\, \tx_{13} \tx_{23}{}^{\!\! -1}\tx_{24} \tx_{14}{}^{\!\! -1} = {}& - u\,
\tx_{13} \tx_{43}{}^{\!\! -1}\tx_{42} \tx_{12}{}^{\!\! -1} + 1 \, , \cr}
}
we find the following relations necessary to 
satisfy the superconformal Ward identity, to cancel terms involving
$\Lambda_{1(23)}{}^{\! 2}\Lambda_{1(24)}\tx_{14} \tx_{24}{}^{\!\! -1} \bep_2$,
\eqn\aa{
\big ( q_4 + v\pr_v + (u-1) \pr_u ) f_{123} 
+ u^{q_1+q_2-2}v^{q_2+q_3-1} \big ( q_2 + v\pr_v + u \pr_u \big ) f_{134}
- u^{q_1+q_2-2}q_1 f_{234} = 0 \, ,
}
to cancel
$\Lambda_{1(23)}{}^{\! 2}\Lambda_{1(24)}\tx_{13} \tx_{23}{}^{\!\! -1} \bep_2$,
\eqn\bb{
\pr_u f_{123} - u^{q_1+q_2-2}v^{q_2+q_3-1}\pr_v f_{134} =0 \, , 
}
to cancel
$\Lambda_{1(24)}{}^{\! 2}\Lambda_{1(23)}\tx_{14} \tx_{24}{}^{\!\! -1} \bep_2$,
\eqn\cc{
\pr_u f_{124} + u^{q_1+q_2-2}\big ( q_2 + v\pr_v + u \pr_u \big ) f_{134} 
= 0 \, ,
}
to cancel
$\Lambda_{1(24)}{}^{\! 2}\Lambda_{1(23)}\tx_{13} \tx_{23}{}^{\!\! -1} \bep_2$,
\eqn\dd{
(\pr_v + \pr_u)f_{124} + u^{q_1+q_2-2} \pr_v f_{134} +
u^{q_1+q_2-2} v^{q_1+q_4-2} q_1 f_{234} = 0 \, ,
}
to cancel
$\Lambda_{1(23)}{}^{\! 2}\Lambda_{1(24)}
\tx_{13} \tx_{43}{}^{\!\! -1}\tx_{42} \tx_{12}{}^{\!\! -1} \bep_1$,
\eqn\ee{
\pr_u f_{123} + u^{q_1+q_2-2}\big ( q_1 + v\pr_v + u \pr_u \big ) f_{234} =0\, ,
}
to cancel
$\Lambda_{1(24)}{}^{\! 2}\Lambda_{1(23)}
\tx_{13} \tx_{43}{}^{\!\! -1}\tx_{42} \tx_{12}{}^{\!\! -1} \bep_1$,
\eqn\ff{
- \pr_u f_{124} + u^{q_1+q_2-2} v^{q_1+q_4-1} \pr_v f_{234} = 0 \, ,
}
to cancel
$\Lambda_{1(23)}{}^{\! 2}\Lambda_{1(24)}\bep_1$,
\eqn\gg{
(v\pr_v + u \pr_u ) f_{123} + u^{q_1+q_2-2}v^{q_2+q_3-1} q_2 f_{134}
- u^{q_1+q_2-2}\big ( (v-u)( q_1 + v\pr_v + u \pr_u ) - v\pr_v \big ) f_{234}
= 0 \, , 
}
and finally to remove $\Lambda_{1(24)}{}^{\! 2}\Lambda_{1(23)}\bep_1$,
\eqn\hh{
\big ( q_3 + v\pr_v + u \pr_u \big ) f_{124} - u^{q_1+q_2-2} q_2 f_{134}
+ u^{q_1+q_2-2} v^{q_1+q_4-1} \big ( q_1 + (v-1)\pr_v + u \pr_u \big )
f_{234} = 0 \, .
}

By taking linear combinations the above eight conditions, \aa$,\dots$,\hh,
may be expressed more succinctly as
\eqna\cond$$\eqalignno{
\pr_u f_{123} = {}& u^{q_1+q_2-2}v^{q_2+q_3-1} \pr_v f_{134} \, , & \cond a \cr
= {}& - u^{q_1+q_2-2} ( q_1 + v\pr_v + u \pr_u ) f_{234} \, , & \cond b \cr
\pr_v f_{123} = {}& u^{q_1+q_2-1} v^{q_2+q_3-2} \pr_u f_{134} \, , & \cond c \cr
\pr_u f_{124} = {}& u^{q_1+q_2-2}v^{q_1+q_4-1}\pr_v f_{234} \, , & \cond d \cr
= {}& - u^{q_1+q_2-2} ( q_2  + v\pr_v + u \pr_u ) f_{134} \, , & \cond e \cr
\pr_v f_{124} = {}& u^{q_1+q_2-1} v^{q_1+q_4-2} \pr_u f_{234} \, , & \cond f \cr
(q_4+ v\pr_v + u \pr_u ) f_{123} = {}& 
- u^{q_1+q_2-1} \pr_u f_{234} \, , & \cond g \cr
(q_3 + v\pr_v + u \pr_u ) f_{124} ={}&
- u^{q_1+q_2-1} \pr_u f_{134} \, . & \cond h \cr}
$$
An important consistency check is that this set of equations are invariant
under simultaneous permutations of $z_{i+}$ and $q_i$. 
For the cyclic permutation $z_{1+}\to z_{2+}\to z_{3+}\to 
z_{4+}\to z_{1+}$, which implies $u\leftrightarrow v$, and also letting
$q_1 \to q_2 \to q_3 \to q_4 \to q_1$ then the equations are
invariant for $f_{123}\to f_{234} \to f_{134}\to f_{124} \to f_{123}$
as expected from the form of \fourp. Similarly under $z_{1+}\leftrightarrow
z_{2+}$, when $u \to u'=u/v, \ v\to v'=1/v$ so that $u\pr_u = u'\pr_{u'}, \
v\pr_v = - v'\pr_{v'} - u'\pr_{u'}$, then also taking $q_1 \leftrightarrow q_2$
and letting $f_{123}(u,v) \to v'{}^{q_4} f_{123}(u',v'), \
f_{124}(u,v) \to v'{}^{q_3} f_{124}(u',v'), \ f_{134}(u,v) \to v'{}^{q_1} 
f_{234}(u',v'), \ f_{234}(u,v) \to v'{}^{q_2} f_{134}(u',v')$, we have
\cond{a}$\,\leftrightarrow\,$\cond{b}, \cond{c}$\,\leftrightarrow\,$\cond{g},
\cond{d}$\,\leftrightarrow\,$\cond{e} and \cond{f}$\,\leftrightarrow\,$\cond{h}.
It is straightfoward to obtain equations for $f_{123}$ etc. alone. Eliminating
$f_{234}$ from \cond{b} and \cond{g} gives
\eqn\fuu{\eqalign{
\big \{& (u^2-u) \pr_u{}^{\! 2} + 2uv \pr_u \pr_v + v^2 \pr_v{}^{\! 2} \cr
{}& + (3-q_2+q_4) ( v\pr_v + u \pr_u ) - (q_3+q_4-1) \pr_u + (2-q_2)q_4 \big \}
f_{123} = 0 \, , \cr}
}
while from \cond{a} and \cond{c} we have
\eqn\fuv{
\big \{ u \pr_u{}^{\! 2} - v \pr_v{}^{\! 2} + (q_3+q_4-1) \pr_u -
(q_1+q_4-1) \pr_v \big \} f_{123} = 0 \, .
}
Similar equations are easily found for $f_{124}, f_{134}, f_{234}$.\foot{To compare
the results here with those of Pickering and West in \Pick\  which are expressed 
in terms of functions
$f,g,k,l$ then $u^{q_3+q_4}v^{q_1+q_4-1}f_{123} = f$, $u^{q_3+q_4}f_{124} = k$,
$f_{134} = l$ and $v^{q_1+q_4-1}f_{234} = g$ and we should let
$q_i \to \thir q_i$. It is difficult to compare our set of eight equations with
theirs although their eqs. (100) and (98) are identical to the equations
corresponding to \fuu\ and \fuv\ for $f_{134}$, and we have verified that their
equations (89,90,91,92) transform appropriately under cyclic permutations 
using the above relations for $f,g,k,l$.}

Eqs. \fuu\ and \fuv\ are identical with a particular generalisation of the
differential equation defining the hypergeometric function to two variables \Erd,
\eqna\hyper$$\eqalignno{
& x(1-x)f_{xx} - y^2 f_{yy} -2xy f_{xy} \cr
{}& + \big ( \gamma - (\alpha+\beta+1)x \big )
f_x - (\alpha+\beta+1)y f_y - \alpha \beta f = 0 \, , & \hyper{a}\cr 
& y(1-y)f_{yy} - x^2 f_{xx} -2xy f_{xy} \cr
{}& + \big ( \gamma' - (\alpha+\beta+1)y \big )
f_y - (\alpha+\beta+1)x f_x - \alpha \beta f = 0 \, , & \hyper{b}\cr}
$$
since \fuu\ has the same form as \hyper{a} and \fuv\ corresponds exactly to the
difference of \hyper{a} and \hyper{b}. These equations have four independent
solutions,
\eqn\hypsol{\eqalign{
{}& F_4(\alpha,\beta,\gamma,\gamma';x,y) \, , \qquad \quad
x^{1-\gamma}F_4(\alpha+1-\gamma,\beta+1-\gamma,2-\gamma,\gamma';x,y) \, , \cr
{}& y^{1-\gamma'}F_4(\alpha+1-\gamma',\beta+1-\gamma',\gamma,2-\gamma';x,y) \, , \cr
{}& x^{1-\gamma} y^{1-\gamma'} F_4(\alpha+2-\gamma-\gamma',\beta+2-\gamma-\gamma',
2-\gamma,2-\gamma';x,y) \, , \cr}
}
with $F_4$, introduced by Appell in 1880\foot{For historical references
see \Ape.}, defined by
\eqn\defF{
F_4(\alpha,\beta,\gamma,\gamma';x,y) = F_4(\beta,\alpha,\gamma,\gamma';x,y)
= \sum_{m,n=0} {(\alpha)_{m+n}(\beta)_{m+n}
\over m! n! \,(\gamma)_m (\gamma')_n} \, x^m y^n \, , 
}
where
\eqn\Poch{
(\gamma)_m = {\Gamma(\gamma+m) \over \Gamma(\gamma)} \, .
}
The series in \defF\ is convergent for $|x|^{1\over 2} + |y|^{1\over 2} <1$. 
The functions $F_4$ have two crucial, for later use, symmetry 
properties,
\eqn\sxy{
F_4(\alpha,\beta,\gamma,\gamma';x,y) = F_4(\alpha,\beta,\gamma',\gamma;y,x) \, ,
}
which follows trivially from \defF, and also, under analytic continuation,
\eqn\sxx{\eqalign{ \!\!\!\!\!\!\!\!\!\!\!
F_4(\alpha,\beta,\gamma,\gamma';x,y) = {}& {\Gamma(\gamma')\Gamma(\beta-\alpha)
\over \Gamma(\gamma'-\alpha) \Gamma(\beta)} \, (-y)^{-\alpha}
F_4(\alpha, \alpha+1-\gamma',\gamma,\alpha+1-\beta;x/y,1/y) \cr
{}+ {}& {\Gamma(\gamma')\Gamma(\alpha-\beta)\over \Gamma(\gamma'-\beta)\Gamma(\alpha)}
\, (-y)^{-\beta}F_4(\beta+1-\gamma',\beta,\gamma,\beta+1-\alpha;x/y,1/y) \, . \cr}
}

With the aid of the solutions of \hyper{a,b} given by \hypsol\ the general
solution of \fuu\ and \fuv\ may then be written as
\eqn\fone{\eqalign{
f_{123}(u,v) = {}&  a\, F_4(q_4,2-q_2,q_3+q_4-1,q_1+q_4-1;u,v) \cr
& {} + b \, u^{q_1+q_2-1}F_4(2-q_3,q_1+1,q_1+q_2,q_1+q_4-1;u,v) \cr
&{} + c\, v^{q_2+q_3-1} F_4(2-q_1,q_3+1,q_3+q_4-1,q_2+q_3;u,v) \cr
&{} + d\, u^{q_1+q_2-1} v^{q_2+q_3-1} F_4(q_2+1,3-q_4,q_1+q_2,q_2+q_3;u,v) \, ,\cr}
}
where we have used \sumq\ extensively. The complete solution of \cond{a,...,h} is
then given by \fone\ along with
\eqnn\ftwo
$$\eqalignno{
f_{124}(u,v) = {}&  - a\, {q_4\over q_1+q_4-1}\,
v^{q_1+q_4-1} F_4(q_4+1,2-q_2,q_3+q_4-1,q_1+q_4;u,v) \cr
& {} - b \, {2-q_3\over q_1+q_4-1}\,  u^{q_1+q_2-1}v^{q_1+q_4-1}
F_4(3-q_3,q_1+1,q_1+q_2,q_1+q_4;u,v) \cr
&{} - c\, {1\over q_3}(q_2+q_3-1) \, F_4(2-q_1,q_3,q_3+q_4-1,q_2+q_3-1;u,v) \cr
&{} - d\, {q_2+q_3-1\over 2-q_4} \, u^{q_1+q_2-1} 
F_4(q_2+1,2-q_4,q_1+q_2,q_2+q_3-1;u,v) \, , & \ftwo \cr}
$$
and 
\eqnn\fthree
$$\eqalignno{
f_{134}(u,v) = {}&   a\, {q_4(2-q_2)\over (q_3+q_4-1)(q_1+q_4-1)}\,
u^{q_3+q_4-1}v^{q_1+q_4-1} F_4(q_4+1,3-q_2,q_3+q_4,q_1+q_4;u,v) \cr
& {} + b \, {q_1+q_2-1\over q_1+q_4-1}\,  v^{q_1+q_4-1}
F_4(2-q_3,q_1+1,q_1+q_2-1,q_1+q_4;u,v) \cr
&{} + c\, {q_2+q_3-1\over q_3+q_4-1} \, u^{q_3+q_4-1}
F_4(2-q_1,q_3+1,q_3+q_4,q_2+q_3-1;u,v) \cr
&{} + d\, {(q_1+q_2-1)(q_2+q_3-1)\over q_2(2-q_4)} \,  
F_4(q_2,2-q_4,q_1+q_2-1,q_2+q_3-1;u,v) \, , & \fthree \cr}
$$
and
\eqnn\ffour
$$\eqalignno{\!\!\!\!\!
f_{234}(u,v) = {}&  - a\, {q_4\over q_3+q_4-1}\,
u^{q_3+q_4-1} F_4(q_4+1,2-q_2,q_3+q_4,q_1+q_4-1;u,v) \cr
& {} - b \, {1\over q_1}(q_1+q_2-1)\,  
F_4(2-q_3,q_1,q_1+q_2-1,q_1+q_4-1;u,v) \cr
&{} - c\, {2-q_1\over q_3+q_4-1} \, u^{q_3+q_4-1}v^{q_2+q_3-1}
F_4(3-q_1,q_3+1,q_3+q_4,q_2+q_3;u,v) \cr
&{} - d\, {q_1+q_2-1\over 2-q_4} \, v^{q_2+q_3-1} 
F_4(q_2+1,2-q_4,q_1+q_2-1,q_2+q_3;u,v) \, . & \ffour \cr}
$$
In each case the $F_4$ functions, defined by \defF, satisfy $\alpha+\beta =
\gamma+\gamma'+1$.\foot{For particular values of $q_i$ the infinite series
for $F_4$ truncates. Thus for $q_2=0$ from the terms $\propto d$ we have
$f_{134}={\rm const.}$ while $f_{123}=f_{124}=f_{234}=0$, which corresponds
to the situation when the four point function reduces to a three point function.
If $q_1=-1$ there is a solution $f_{123}=b u^{q_2-2}$, 
$f_{124} = b{q_3-2 \over q_4-2} \, u^{q_2-2} v^{q_4-2}$,
$f_{134} = b{q_2-2 \over q_4-2} \, v^{q_4-2}$, $f_{234} = b(q_2-2) ( 1
+ {q_3-2\over q_2-2} u + {q_3-2\over q_4-2}v )$. For $q_2=0$ as well this
coincides, up to misprints, with the solution (104) in \Pick.}

\newsec{Crossing Symmetry Relations}

Although the above results provide a complete solution of the superconformal
Ward identities, the solutions of the differential equations are further
constrained by considering permutations of $z_{i+}$ and simultaneously $q_i$.
Such permutations act on the invariants $u,v$ so that in general they are
not restricted to the domain of convergence of the $F_4$ functions and thus
analytic continuation is necessary. It is essential that the results be
independent of the path of analytic continuation or that they be monodromy 
invariant.

Firstly for cyclic permutations, when $u\leftrightarrow v$, in \fourp\
we require
\eqn\permc{\eqalign{
f_{123}(& q_1,q_2,q_3,q_4;u,v) = f_{124}(q_2,q_3,q_4,q_1;v,u) \cr
& = f_{134}(q_3,q_4,q_1,q_2;u,v) = f_{234}(q_4,q_1,q_2,q_3;v,u) \, ,\cr}
}
which, with \sxy, leads to
\eqna\pabcd$$
\eqalignno{
a(q_1,q_2,q_3,q_4) = {}& -{1\over q_4}(q_3+q_4-1) c(q_2,q_3,q_4,q_1) \, , 
& \pabcd{a}\cr
d(q_1,q_2,q_3,q_4) = {}& - {2-q_4 \over q_1+q_2-1} \, b(q_2,q_3,q_4,q_1) \,,
& \pabcd{b} \cr
b(q_1,q_2,q_3,q_4) = {}& - {q_1 \over q_1+q_2-1} \, a(q_2,q_3,q_4,q_1) \, , 
& \pabcd{c} \cr
c(q_1,q_2,q_3,q_4) = {}& - {q_3+q_4-1\over 2-q_1} \, d(q_2,q_3,q_4,q_1) \, . 
& \pabcd{d} \cr}
$$

The critical conditions arise from considering 
$z_{1+},q_1 \leftrightarrow z_{2+},q_2$ so that $u\to u'=u/v, \, v \to v'=1/v$.
In this case we require
\eqna\permb
$$\eqalignno{
f_{123}(& q_1,q_2,q_3,q_4;u,v) = v'{}^{q_4} f_{123}(q_2,q_1,q_3,q_4;u',v') \, ,
& \permb{a} \cr
f_{124}(& q_1,q_2,q_3,q_4;u,v) = v'{}^{q_3} f_{124}(q_2,q_1,q_3,q_4;u',v') \, , 
& \permb{b} \cr
f_{134}(& q_1,q_2,q_3,q_4;u,v) = v'{}^{q_1} f_{234}(q_2,q_1,q_3,q_4;u',v') \, ,
& \permb{c} \cr}
$$
Using now \sxx, we get from \permb{a}, for one path of
analytic continuation,
\eqna\pab
$$\eqalignno{
\!\!\!\!\!\!\!\!\!\!\!\!\!\!\!
\bigg \{ a(q_1,q_2,q_3,q_4&){e^{-i\pi q_4}\Gamma(q_1+q_4-1)\over \Gamma(q_1-1)
\Gamma(2-q_2)} 
+c (q_1,q_2,q_3,q_4){e^{i\pi (q_1-2)}\Gamma(q_2+q_3)\over \Gamma(1-q_4)
\Gamma(q_3+1)}\bigg \} \Gamma(q_1+q_3-1) \cr
&{} = a(q_2,q_1,q_3,q_4) \, , &\pab{a}\cr
\!\!\!\!\!\!\!\!\!\!\!\!\!\!\!\!\!
\bigg \{  a(q_1,q_2,q_3,q_4)&{e^{i\pi (q_2-1)}\Gamma(q_1+q_4-1)\over \Gamma(-q_3)
\Gamma(q_4)} 
+c (q_1,q_2,q_3,q_4){e^{-i\pi (q_3+1)}\Gamma(q_2+q_3)\over \Gamma(q_2-1)
\Gamma(2-q_1)}\bigg \} \Gamma(q_2+q_4-2) \cr
&{} = c(q_2,q_1,q_3,q_4) \, , &\pab{b}\cr
\!\!\!\!\!\!\!\!\!\!\!\!\!\!\!\!\!
\bigg \{  b(q_1,q_2,q_3,q_4)&{e^{i\pi (q_3-2)}\Gamma(q_1+q_4-1)\over \Gamma(-q_2)
\Gamma(q_1+1)} 
+d (q_1,q_2,q_3,q_4){e^{-i\pi (q_2+1)}\Gamma(q_2+q_3)\over \Gamma(q_3-1)
\Gamma(3-q_4)}\bigg \} \Gamma(q_1+q_3-1) \cr
&{} = b(q_2,q_1,q_3,q_4) \, , &\pab{c}\cr
\!\!\!\!\!\!\!\!\!\!\!\!\!\!\!\!\!
\bigg \{  b(q_1,q_2,q_3,q_4)&{e^{-\pi (q_1+1)}\Gamma(q_1+q_4-1)\over \Gamma(q_4-2)
\Gamma(2-q_3)} 
+d (q_1,q_2,q_3,q_4){e^{i\pi (q_4-3)}\Gamma(q_2+q_3)\over \Gamma(-q_1)
\Gamma(q_2+1)}\bigg \} \Gamma(q_2+q_4-2) \cr
&{} = d(q_2,q_1,q_3,q_4) \, . & \pab{d}\cr}
$$
For consistency the imaginary parts of the left hand sides of \pab{a,..,d}
must vanish. In this case the same result is obtained for the other possible
analytic continuation since their difference is just in the imaginary
part.\foot{For a related discussion
in two dimensions see \CFT.} From \pab{a} or \pab{c}, using $\Gamma(x)\Gamma(1-x)
= - \Gamma(x+1)\Gamma(-x) = \pi/\sin \pi x$, we get
\eqn\relac{
c(q_1,q_2,q_3,q_4) = - {\Gamma(q_1+q_4-1)\Gamma(2-q_1)\Gamma(q_3+1) \over
\Gamma(q_2+q_3) \Gamma(2- q_2) \Gamma(q_4) }\, a(q_1,q_2,q_3,q_4) \, ,
}
while from \pab{c} or \pab{d} we also obtain
\eqn\relbd{
d(q_1,q_2,q_3,q_4) = - {\Gamma(q_1+q_4-1)\Gamma(3-q_4)\Gamma(q_1+1) \over
\Gamma(q_2+q_3) \Gamma(2- q_3) \Gamma(q_1+1) }\, b(q_1,q_2,q_3,q_4)\, .
}
Using \relac, \pab{a} reduces to
\eqn\saa{
a(q_1,q_2,q_3,q_4)\, {\Gamma(q_1+q_3-1)\Gamma(2-q_1)\over
\Gamma(q_2+q_3-1)\Gamma(2-q_2)} = a(q_2,q_1,q_3,q_4)\, ,
}
while \pab{b} gives an equivalent, by virtue of \relac, relation for $c$.
Similarly \pab{c} or \pab{d} give
\eqn\sbb{
b(q_1,q_2,q_3,q_4)\, {\Gamma(q_1+q_3-1)\Gamma(q_2+1)\over
\Gamma(q_2+q_3-1)\Gamma(q_1+1)} = b(q_2,q_1,q_3,q_4)\, .
} 
Identical results may also be obtained from \permb{b} or \permb{c}.

To simplify further, it is convenient to define
\eqn\defA{\eqalign{
a(q_1,q_2,q_3,q_4) = {}& \Gamma(q_2+q_3-1)\Gamma(q_1+q_2-1)\Gamma(2-q_2)
\Gamma(q_4) \, A(q_1,q_2,q_3,q_4) \, , \cr
c(q_1,q_2,q_3,q_4) = {}& \Gamma(q_1+q_4-2)\Gamma(q_1+q_2-1)\Gamma(2-q_1)
\Gamma(q_3+1) \, A(q_1,q_2,q_3,q_4) \, , \cr}
}
and also
\eqn\defB{\eqalign{
b(q_1,q_2,q_3,q_4) = {}& \Gamma(q_2+q_3-1)\Gamma(q_3+q_4-2)\Gamma(2-q_3)
\Gamma(q_1+1) \, B(q_1,q_2,q_3,q_4) \, , \cr
d(q_1,q_2,q_3,q_4) = {}& \Gamma(q_1+q_4-2)\Gamma(q_3+q_4-2)\Gamma(3-q_4)
\Gamma(q_2+1) \, B(q_1,q_2,q_3,q_4) \, , \cr}
}
where the second lines follow from \relac\ and \relbd\ respectively. Now \saa\
and \sbb\ become
\eqn\sAB{
A(q_1,q_2,q_3,q_4) = A(q_2,q_1,q_3,q_4) \, , \qquad
B(q_1,q_2,q_3,q_4) = B(q_2,q_1,q_3,q_4) \, .
}
Inserting \defA\ into \pabcd{a} and \defB\ into \pabcd{b} gives
\eqn\cAA{
A(q_1,q_2,q_3,q_4) = A(q_2,q_3,q_4,q_1) \, , \qquad
B(q_1,q_2,q_3,q_4) = B(q_2,q_3,q_4,q_1) \, .
}
Similarly \pabcd{c} and \pabcd{d} lead to
\eqn\cAB{
B(q_1,q_2,q_3,q_4) = A(q_2,q_3,q_4,q_1) \, , \qquad
A(q_1,q_2,q_3,q_4) = B(q_2,q_3,q_4,q_1) \, .
}
In consequence we must have $B=A$ so that the chiral four point function is
determined up to a single overall constant. $f_{123}(u,v)$ is given by
substituting \defA\ and \defB, with $A=B$, into \fone\ while $f_{124}(u,v),
\, f_{134}(u,v), \, f_{234}(u,v)$ are given by similar expressions which may
easily be found from \permc.

For completeness we may also consider the permutation $z_{1+},q_1
\leftrightarrow z_{3+},q_3$ and $z_{2+},q_2\leftrightarrow z_{4+},q_4$,
when $u,v$ are invariant. This requires $f_{123}(q_3,q_4,q_1,q_2;u,v)
= f_{134}(q_1,q_2,q_3,q_4;u,v)$ and $f_{124}(q_3,q_4,q_1,q_2;u,v)
= f_{234}(q_1,q_2,q_3,q_4;u,v)$. This relates $a$ to $d$ and $b$ to $d$
and with \defA\ and \defB\ would require $A(q_3,q_4,q_1,q_2)=B(q_1,q_2,q_3,q_4)$
and $B(q_3,q_4,q_1,q_2)=A(q_1,q_2,q_3,q_4)$.

\newsec{Short Distance Expansions}

In the limits  $x_{1+} \to x_{2+}$ or $x_{3+} \to x_{4+}$ we have, with
the definitions in \defuv, $u\to 0, \, v\to 1$. Similarly for
$x_{1+} \to x_{4+}$ or $x_{2+} \to x_{3+}$, $u\to 1, \, v\to 0$ while for
$x_{1+} \to x_{3+}$ or $x_{2+} \to x_{4+}$, $1/v\to 0, \, u/v \to 1$. In order
to understand the behaviour in these short distance limits a different
expansion, which reveals the form of the $F_4$
function for one of the arguments near the singular point 1, 
than that given by \defF\ is necessary.

In the simpler case of standard hypergeometric functions the relevant results
are easily obtained. The associated second order ordinary differential equation 
has three singular points at $0, 1, \infty$. The two independent solutions
may be taken as $F(\alpha,\beta;\gamma;x)$ and
$x^{1-\gamma}F(\alpha+1-\gamma,\beta+1-\gamma;2-\gamma;x)$, which are 
thus given as an expansion in powers of $x$, 
but they can be equally expressed in
terms of  $F(\alpha,\beta;\alpha+\beta + 1 - \gamma;1-x)$ and
$(1-x)^{\gamma-\alpha -\beta}
F(\gamma-\alpha,\gamma-\beta;\gamma-\alpha -\beta+1;1-x)$, determining
the form as $x\to 1$, or $(-x)^{-\alpha}
F(\alpha,\alpha+1-\gamma;\alpha+1-\beta;1/x)$ and $(-x)^{-\beta}
F(\beta+1-\gamma,\beta;\beta+1-\alpha;1/x)$, which gives an equivalent expression
in terms of $1/x$, revealing the behaviour at $\infty$. Any of these functions
may be written in terms of a linear combination of the two functions defined by
a series expansion at either of the other singular points.
For the $F_4$ function, given by \defF, \sxx\ gives an equivalent result 
involving the behaviour for one variable at $\infty$.
To determine an analogous result for the form of $F_4$, as used here, for
one argument approaching $1$ we define the new function
\eqn\defG{
G(\alpha,\beta,\gamma,\delta;x,y) = \sum_{m,n=0} {(\delta-\alpha)_m
(\delta-\beta)_m \over m! (\gamma)_m} \, 
{(\alpha)_{m+n} (\beta)_{m+n} \over n! (\delta)_{2m+n}} \, x^m y^n \, .
}
Writing the sum over $n$ in terms of a hypergeometric function and using
the identity relating hypergeometric functions of arguments $y$ and $1-y$
we may obtain\foot{The behaviour of $F_4$ functions
near $y=1$ was investigated quite recently by Exton \Ext. The function $G$
is essentially that defined by Exton and the relation obtained here is
a special case of his formulae. The series for $G$ in \defG\ also features
in a related discussion in \Lang.}
\eqn\PF{\eqalign{
G(\alpha,\beta,\gamma & ,\delta;x,1-y) = 
{\Gamma(\delta) \Gamma(\delta-\alpha-\beta)\over \Gamma(\delta-\alpha)
\Gamma(\delta-\beta)} \, F_4(\alpha,\beta,\gamma,\alpha+\beta+1-\delta;x,y) \cr
{}& + {\Gamma(\delta) \Gamma(\alpha+\beta-\delta)\over \Gamma(\alpha)
\Gamma(\beta)} \, y^{\delta-\alpha-\beta}
F_4(\delta-\alpha,\delta-\beta,\gamma,\delta-\alpha-\beta+1;x,y) \,.\cr}
}
{}From this it is evident that
\eqn\sP{
G(\alpha,\beta,\gamma ,\delta;x,1-y) = y^{\delta-\alpha-\beta}
G\big (\delta-\alpha,\delta-\beta,\gamma ,\delta;x,1-y\big ) \, ,
}
and using \sxx\ and standard $\Gamma$-function identities we find, from \PF,
\eqn\sPP{
G(\alpha,\beta,\gamma ,\delta;x,1-y) = y^{-\alpha}
G\big (\alpha,\delta-\beta,\gamma ,\delta;x/y,1-1/y\big ) \, .
}

Using \PF\ with \fone\ where $a,b,c,d$ are given by \defA\ and \defB\ we may
now find in terms of the function $G$,
\eqn\fph{\eqalign{
f_{123}(u,v) = {}&{ \Gamma(q_1+q_2-1)\Gamma(2-q_1)\Gamma(2-q_2)\Gamma(q_3+1)
\Gamma(q_4)\over \Gamma(q_3+q_4+1) } \, A \cr
{}&\times  G(q_4,2-q_2,q_3+q_4-1, q_3+q_4+1;u,1-v) \cr
{}& + { \Gamma(q_3+q_4-2)\Gamma(q_1+1)\Gamma(q_2+1)\Gamma(2-q_3)
\Gamma(3-q_4)\over \Gamma(q_1+q_2+2) } \, B \cr
{}&\times  u^{q_1+q_2-1}G(2-q_3,q_1+1,q_1+q_2, q_1+q_2+2;u,1-v) \, , \cr}
}
and
\eqn\fphb{\eqalign{
f_{124}(u,v) = {}&{ \Gamma(q_1+q_2-1)\Gamma(2-q_1)\Gamma(2-q_2)\Gamma(q_3)
\Gamma(q_4+1)\over \Gamma(q_3+q_4+1) } \, A \cr
{}&\times  G(q_3,2-q_1,q_3+q_4-1, q_3+q_4+1;u,1-v) \cr
{}& + { \Gamma(q_3+q_4-2)\Gamma(q_1+1)\Gamma(q_2+1)\Gamma(3-q_3)
\Gamma(2-q_4)\over \Gamma(q_1+q_2+2) } \, B \cr
{}&\times  u^{q_1+q_2-1}G(2-q_4,q_2+1,q_1+q_2, q_1+q_2+2;u,1-v) \, , \cr}
}
and
\eqn\fphc{\eqalign{
f_{134}(u,v) = {}&
{ \Gamma(q_3+q_4-1)\Gamma(q_1+1)\Gamma(q_2)\Gamma(2-q_3)
\Gamma(2-q_4)\over \Gamma(q_1+q_2+1) } \, B \cr
{}&\times  G(q_2,2-q_4,q_1+q_2-1, q_1+q_2+1;u,1-v) \cr
{}& + { \Gamma(q_1+q_2-2)\Gamma(2-q_1)\Gamma(3-q_2)\Gamma(q_3+1)
\Gamma(q_4+1)\over \Gamma(q_3+q_4+2) } \, A \cr
{}&\times  u^{q_3+q_4-1}G(2-q_1,q_3+1,q_3+q_4, q_3+q_4+2;u,1-v) \, , \cr}
}
and
\eqn\fphd{\eqalign{
f_{234}(u,v) = {}&
{ \Gamma(q_3+q_4-1)\Gamma(q_1)\Gamma(q_2+1)\Gamma(2-q_3)
\Gamma(2-q_4)\over \Gamma(q_1+q_2+1) } \, B \cr
{}&\times  G(q_1,2-q_3,q_1+q_2-1, q_1+q_2+1;u,1-v) \cr
{}& + { \Gamma(q_1+q_2-2)\Gamma(3-q_1)\Gamma(2-q_2)\Gamma(q_3+1)
\Gamma(q_4+1)\over \Gamma(q_3+q_4+2) } \, A \cr
{}&\times  u^{q_3+q_4-1}G(2-q_2,q_4+1,q_3+q_4, q_3+q_4+2;u,1-v) \, . \cr}
}
With these expressions the relations \permb{a,b,c} are easy to verify using
\sP\ and \sPP. Clearly the results are manifestly analytic at $v=1$.

In general if we define
\eqn\defH{\eqalign{
H(&\alpha,\beta,\gamma,\delta;u,v) = {\Gamma(1-\gamma)\over \Gamma(\delta)}
\Gamma(\alpha) \Gamma(\beta) \Gamma(\delta-\alpha) \Gamma(\delta-\beta) \,
G(\alpha,\beta,\gamma,\delta;u,1-v) \cr
{}& + {\Gamma(\gamma-1)\over \Gamma(\delta-2\gamma+2)}
\Gamma(\alpha-\gamma+1) \Gamma(\beta -\gamma+1) \Gamma(\delta-\gamma-\alpha+1) 
\Gamma(\delta-\gamma-\beta+1) \cr
& \qquad\qquad\qquad{}\times u^{1-\gamma}G(\alpha-\gamma+1, \beta -\gamma+1, 
2-\gamma, \delta-2\gamma+2;u,1-v) \, , }
}
then, from \PF\ and \sxy, we have
\eqn\sHH{
H(\alpha,\beta,\gamma,\delta;u,v) = H(\alpha,\beta,\alpha+\beta+1-\delta,
\alpha+\beta+1-\gamma;v,u) \, .
}
The expression given by \defH\ coincides, for appropriate choices of
$\alpha,\beta,\gamma,\delta$, with the forms in 
\fph,\fphb,\fphc,\fphd\ if $A=B$. From \sPP\ we have
\eqn\sH{
H(\alpha,\beta,\gamma ,\delta;u,v) =  v^{-\alpha}
H\big (\alpha,\delta-\beta,\gamma ,\delta;u/v,1/v\big ) \, ,
}
and from \sP\ together with \sHH\ we may obtain
\eqn\sHHH{\eqalign{
H(\alpha,\beta,\gamma,\delta &;u,v) =  v^{\de-\alpha-\beta}
H\big (\de-\alpha,\delta-\beta,\gamma ,\delta;u,v\big ) \cr
= {}& u^{1-\gamma} v^{\de-\alpha-\beta}
H\big (\de-\alpha+1-\gamma,\delta-\beta+1-\gamma,2-\gamma ,
\delta-2\gamma+2;u,v\big ) \, . \cr}
}
Together \sHH, \sH\ and \sHHH\
are sufficient to obtain the necessary symmetry relations. 

If $\gamma=1$ the definition \defH\ reduces to
\eqn\defHH{\eqalign{
\!\!\!\!\!\!\!\!\!\! H(\alpha,\beta,1,\delta;u,v) ={}&  {1\over \Gamma(\delta)}
\Gamma(\alpha) \Gamma(\beta) \Gamma(\delta-\alpha) \Gamma(\delta-\beta) \,
\bigg \{\! - \ln u \, G(\alpha,\beta,1,\delta;u,1-v) \cr
{}& + \sum_{m,n=0} {(\delta-\alpha)_m
(\delta-\beta)_m \over (m!)^2} \,
{(\alpha)_{m+n} (\beta)_{m+n} \over n! (\delta)_{2m+n}} \, f_{mn} \, 
u^m (1-v)^n \bigg \} \, , \cr
f_{mn} = {}& 2\psi(1+m) + 2\psi(\de+2m+n)
- \psi(\de-\alpha+m) -\psi(\de-\beta+m) \cr
{}& - \psi(\alpha+m+n) - \psi(\beta+m+n) \, , \cr}
}
involving $\ln u$ although \sHH, \sH\ and \sHHH\ remain valid.

\newsec{Counting for higher $N$-point Functions}

The constraints arising from superconformal invariance uniquely determine
the form of the scalar chiral superfield four point function, as shown in
section 3. We attempt here to count the number of independent functions
of the appropriate generalisations of $u,v$ and to determine the number of
independent differential equations constraining them for higher point
functions of scalar chiral superfields.

In order to undertake such an analysis the superconformal group is restricted
by first using superconformal transformations to set
\eqn\ot{
x_{1+} = 0 \, , \quad x_{2+} = \infty \, , \quad \theta_1 =0 \, , 
\quad \theta_2 = 0 \, .
}
The residual symmetry group is then that generated by matrices of the form \MM\
with $a,b,\epsilon,\bta$ zero. The associated subgroup of $G_0$ is given by
rotations and scale, $U(1)_R$, transformations for which the infinitesimal
parameters are $\omega^{ab},\kappa,\bkap$. With \ot\ we have
\eqn\Lr{
\Lambda_{1(2i)} = \theta_i \tx_i{}^{\! -1} \, .
}
After imposing \ot\ an $N$-point function for chiral superfields depends only
on $z_{i+} = (x_{i+}, \theta_i)$, $ i = 3, \dots N$. The rotational and scale
invariant variables constructed from $z_{i+}$ are then easily seen to be
\eqn\cinv{
u_{ij} = {x_{j+}{\cdot x_{i+}}\over x_{3+}{}^{\! 2}} \, ,\ \ j\ge i > 3 \, ,
\qquad u_{i} = {2x_{i+}{\cdot x_{3+}}\over x_{3+}{}^{\! 2}} \, , \ \ i> 3 \, .
}
Evidently there are $\half(N-3)(N-2)$ such $u_{ij}$ and $N-3$ $u_i$ giving
$\half N (N-3)$ in total. However, since we are concerned only with four
dimensional space, if $N\ge 7$ we may express $x_{i+}, \, i\ge 7$ in terms
of $x_{j+}, \, j=3,4,5,6$, with coefficients involving $x_{i+}{\cdot x_{j+}}$. 
This ensures that for the $u_{ij}$ in \cinv\ we may therefore
restrict to $j=3,4,5,6$ and the number of such independent invariants becomes
$N-3 + N-4 + N-5$ giving, with the $u_i$ as well, altogether $4N-15$. 
Clearly for $N\ge 7$ the 15
parameter conformal group $SU(2,2)$ is acting transitively. In fact this
result gives the correct number also for $N=5,6$ whereas for $N=4$ we have
just two invariants $u_{44}, u_4$.

For application to $N$-point functions for chiral superfields with $\sum_i q_i
=3$ we need a set of nilpotent monomials ${\rm O}(\theta^2)$, generalising
$\Lambda_{i(jk)}{}^{\! 2}$, $i<j<k$, used for $N=4$. With the choice in \ot\
we may take as a linearly independent basis
\eqn\dXi{
\Xi_i = \theta_i{}^{\! 2} \, , \ \ i\ge 3 \, , \qquad
\Xi_{ij,r} = \theta_j M_r \tth_i \, , \ \ j>i\ge 3 \, ,
}
where $(M_r)_\alpha{}^{\!\beta}$ are linearly independent $2\times 2$ matrices 
constructed from $x_{i+}$, which we assume to have dimension zero. To achieve
the form in \dXi\ we may note that $\theta_i M \tth_j = \theta_j {\tilde M}
\tth_i$ where ${\tilde M}_\alpha{}^{\!\beta} = \ep_{\alpha\gamma}
\ep^{\beta\de} M_\de{}^{\!\gamma}$. For $N\ge 5$ 
it is sufficient to take just $r=1,2,3,4$ since any such matrix may be expressed
in terms of the basis formed by $1, \, \x_{4+} \tx_{3+}/ x_{3+}{}^{\! 2}, \,
\x_{5+} \tx_{3+}/ x_{3+}{}^{\! 2}, \, \x_{5+} \tx_{4+}/ x_{3+}{}^{\! 2}$. Clearly
there are $N-2$ independent $\Xi_i$ and $2(N-3)(N-2)$ independent $\Xi_{ij,r}$ giving
$(2N-5)(N-2)$ independent such $\Xi_I$ in total. The general $N$-point function can
then be written in terms of scalar functions of the invariants $u$ as
\eqn\Np{
\big (x_{2+}{}^{\! 2}\big )^{q_2}
\big \l \phi_1(z_{1+}) \phi_2  (z_{2+}) \dots \phi_N(z_{N+}) \big \r 
\Big |_{{x_{1+}=0, \theta_1 = 0 \atop x_{2+}\to \infty, \theta_2 = 0}}
= \big (x_{3+}{}^{\! 2}\big )^{q_2 -2} \sum_I \Xi_I \, f_I(u) \, .
}
For $N=4$ there are only two independent matrices $M_r$ since they can be 
restricted to the basis formed by $1, \, \x_{4+} \tx_{3+}/ x_{3+}{}^{\! 2}$.
There are therefore just four independent $\Xi_I$, which can be alternatively
expressed in terms $\Lambda_{1(23)}{}^{\! 2},\Lambda_{1(24)}{}^{\! 2},
\Lambda_{1(34)}{}^{\! 2},\Lambda_{2(34)}{}^{\! 2}$. Thus these results are
in accord with the treatment in section 3 and the expansion of the four point 
function exhibited in \fourp. 
For $N>4$ it is easy to see that the $\Xi_I$ cannot all be expressed solely 
in terms of $\Lambda_{i(jk)}{}^{\! 2}$, for $N=5$ there are 10
$\Lambda_{i(jk)}{}^{\! 2}$, $i<j<k$, whereas the basis in \dXi\ gives 15.

The non trivial constraints in the superconformal Ward identities \scf\ arise
from the terms involving $\bep$ or $\eta$. From \sol\ such terms are
${\rm O}(\theta^3)$, and the number of independent conditions, which involve
linear first order partial differential equations for the $f_I(u)$, is equal
to the number of independent monomials of the form $\theta^3\bep$ or
$\theta^3 \eta$. For $N\ge 4$ we have
\eqn\ttt{
\theta_j{}^{\! 2} \, \theta_i N_s \bep \, , \qquad \theta_j{}^{\! 2} \, \theta_i 
M_r \eta \, , \ \ \ i\ne j \, , \ i,j\ge 3 \, ,
}
where $(N_s)_{\alpha\dal}$ is a linearly independent set of $2\times 2$ matrices. 
For $N\ge 5$ $M_r$ may be reduced to the basis described above and for $N\ge 6$
$N_s$ may be given in terms of the basis  $\x_3,\x_4,\x_5,\x_6$. Thus if $N\ge 6$
there are four possible $M_r$ and also four $N_s$ so that the number of
independent monomials of the form \ttt\ is $8(N-3)(N-2)$. In addition to \ttt\
for $N\ge 5$ we may also construct
\eqn\tttt{
\theta_k M_r \tth_j \, \theta_i N_s \bep \, , \qquad
\theta_k M_r \tth_j \, \theta_i M_s \eta \, , \ \ \ k>j>i\ge 3 \, .
}
The ordering $k>j>i$ is achieved by Fierz type identities, since if $i>j$ we may 
write $\tth_j \theta_i  = - \sum_r \theta_i M_r \tth_j {M'}_{\! r}$ for $\{M_r\}$
forming a basis for $2\times 2$ matrices and where ${M'}_{\! r}$
satisfy ${\rm tr}({M'}_{\! r} M_s ) = \de_{rs}$. The expressions \ttt\ and \tttt\
are a linearly independent basis. Assuming both four linearly independent
$M_r$ and $N_s$ we have ${16\over 3}(N-4)(N-3)(N-2)$ monomials of the form \tttt.
Combining these with those exhibited in  \ttt\ we have ${{8\over 3}(N-3)(N-2)
(2N-5)}$ in total giving this number of constraints on the $f_I$. 
Since there are $(N-2)(2N-5)$ $f_I$ depending on $4N-15$ variables the number of 
functional degrees of freedom remaining after imposing the
differential constraints is ${1\over 3}(N-2)(2N-5)(4N-21)$, which for $N=6$ is 28.

When $N=4$ the relevant monomials are only of the form in \ttt, with $i,j=3,4$, and we
further restrict to $r=1,2$, corresponding to $1, \x_{4+} \tx_{3+}/x_{3+}{}^{\! 2}$,
and also $s=1,2$, since $N_s$ is also restricted to the basis $\x_3,\x_4$. This gives
8 independent monomials so that there are 8 conditions on the 4 functions of 2
variables $f_I$. In appendix A we explicitly obtain the 8 relations for 
$f_1,f_2,f_3,f_4$ which are equivalent to those found earlier in \aa$,\dots,\hh$ 
or \cond{a,..h}.  For $N=4$ we therefore expect a unique functional form for the
solution, up to choices of integration constants, which of course is in accord
with the results of section 3. 
For $N=5$ there are apparently 80 constraints on 15 functions $f_I$ of 5
variables leading to an overdetermined system. However if any $q_i=0$ the equations
should reduce to those determining the four point function so that there
are possible linear dependencies in the constraints but this case is too
complicated to investigate in detail.

\newsec{Operator Product Expansion}

In conformal field theories there are further conditions to be obtained
by imposing the operator product expansion. A four point function involving
fields at $x_1,x_2,x_3,x_4$ may be expanded as a convergent series for
$x_1\approx x_2, \ x_3 \approx x_4$ in terms of an infinite set of 
quasi-primary fields and also equivalently for 
$x_1\approx x_3, \ x_2 \approx x_4$. The equality of the two expansions
provides a constraint on the operator content of the theory. In this section 
we consider the contribution of a single chiral quasi-primary superfield
and its descendents to the four point function of chiral superfields.

The operator product coefficients are determined by the two and three point
functions. For scalar fields $\varphi_i$, with dimension $\de_i$ in 
dimension $d$, then the two point functions in a conformal theory may be
chosen as
\eqn\Twop{
\l \varphi_i (x_1) \varphi_j (x_2) \r = {\de_{ij}\over 
r_{12}{}^{\raise 2pt\hbox{$\scriptstyle \! \de_i$}}} \,,
}
and the three point functions are then
\eqn\Threep{
\l \varphi_i (x_1) \varphi_j (x_2) \varphi_k(x_3) \r =
{C_{ijk} \over r_{12}{}^{\raise 2pt\hbox{$\scriptstyle  {1\over2}
(\de_i + \de_j - \de_k)$}}\,
r_{23}{}^{\raise 2pt\hbox{$\scriptstyle  \! {1\over2}
(\de_j + \de_k - \de_i)$}}\,
r_{31}{}^{\raise 2pt\hbox{$\scriptstyle \! {1\over2}
(\de_k + \de_i - \de_j)$}}} \, .
}
The contribution of the field $\varphi_k$ to the operator product of
$\varphi_i$ and $\varphi_j$ is then determined from \Twop\ and \Threep\ to be
\eqn\OPE{
\varphi_i (x_1) \varphi_j (x_2) \sim C_{ijk} \, 
r_{12}{}^{\! -{1\over2}(\de_i + \de_j - \de_k)}
C^{{1\over2}(\de_k + \de_i - \de_j),{1\over2}(\de_k - \de_i +\de_j)}_
{\de_k + 1- {1\over 2}d}(x_{12}, \pr_{x_2}) \varphi_k(x_2) \, ,
}
where
\eqn\defC{
C^{a,b}_{S+1 - {1\over 2}d}(x_{12},\pr_{x_2})
{1\over r_{23}{}^{\raise 2pt\hbox{$\scriptstyle  \! S$}}} =
{1\over r_{13}{}^{\raise 2pt\hbox{$\scriptstyle \! a$}}
r_{23}{}^{\raise 2pt\hbox{$\scriptstyle \! b$}}} \, , \qquad S=a+b \, .
}
$C^{a,b}_{\kappa}(s,\pr)$ is given as an infinite series in $s{\cdot \pr}$
and $s^2 \pr^2$ in appendix D, $C^{a,b}_{\kappa}(0,\pr) = 1$. For
application to four point functions we have
\eqn\OPEF{\eqalign{
C^{a,b}_{S+1 - {1\over 2}d}(x_{12},\pr_{x_2})
{1\over r_{23}{}^{\raise 2pt\hbox{$\scriptstyle \! e$}}
r_{24}{}^{\raise 2pt\hbox{$\scriptstyle \! f$}}}
={}& {1\over r_{14}{}^{\raise 2pt\hbox{$\scriptstyle \! a$}}
r_{24}{}^{\raise 2pt\hbox{$\scriptstyle \! b$}}} \bigg ({r_{14}\over r_{13}}
\bigg )^{\! e} G(b,e,S+1 - \half d, S; u,1-v) \cr
={}& {1\over r_{13}{}^{\raise 2pt\hbox{$\scriptstyle \! a$}}
r_{23}{}^{\raise 2pt\hbox{$\scriptstyle \! b$}}} \bigg ({r_{13}\over r_{14}}
\bigg )^{\! f} G(a,f,S+1 - \half d, S; u,1-v) \, ,}
}
with $S=a+b=e+f$ and using \sP.

Similar results may be obtained for chiral scalar superfields, restricting 
now to $d=4$. The basic two point function involves a chiral superfield 
$\phi$ and its anti-chiral conjugate $\bphi$ \HO,
\eqn\TwoP{
\l \phi_i(z_{1+}) \bphi_j (z_{2-}) \r = {\de_{ij}\over\big (x_{1+} -
2i \theta_1 \si \bth_2 - x_{2-}\big )^{2q_i}}\,,}
and the corresponding three point function has the form
\eqn\ThreeP{
\l \phi_i(z_{1+}) \phi_j(z_{2+}) \bphi_k (z_{3-}) \r =
{C_{ij{\bar k}} \over\big (x_{1+} - 2i \theta_1 \si \bth_3 - x_{3-}
\big )^{2q_i}
\big (x_{2+} - 2i \theta_2 \si \bth_3 - x_{3-} \big )^{2q_j}} \, ,
}
and is only possible if $q_k=q_i+q_j$. As a consequence of \TwoP\ and
\ThreeP\ the chiral superfield contributes to the operator product
expansion of $\phi_i$ and $\phi_j$
\eqn\OPEC{
\phi_i(z_{1+}) \phi_j(z_{2+}) \sim C_{ij{\bar k}} \, \C^{q_1,q_2}
(z_{12+}, \pr_{z_{2+}}) \phi_k (z_{2+}) \, ,
}
where
\eqn\OPES{\eqalign{
\C^{q_1,q_2}&(z_{12+}, \pr_{z_{2+}})
{1\over\big (x_{2+} - 2i \theta_2 \si \bth - x_{-}\big )^{2q}}\cr
={}& {1 \over\big (x_{1+} - 2i \theta_1 \si \bth - x_{-} \big )^{2q_1}
\big (x_{2+}- 2i\theta_2 \si\bth - x_{-} \big )^{2q_2}}\,,\quad q=q_1+q_2 \, .}
}
As shown in Appendix E we may express $\C^{q_1,q_2}$ as
\eqn\CCC{\eqalign{
\C^{q_1,q_2}(z_{12+}, \pr_{z_{2+}}) ={}& C^{q_1,q_2}_{q-1}
(x_{12+}, \pr_{x_{2+}})
+ {q_1\over q}\,  C^{q_1+1,q_2}_{q-1}
(x_{12+}, \pr_{x_{2+}}) \theta_{12}\pr_{\theta_2} \cr
{}& + {q_1 q_2 \over 2q(q^2-1)}\, C^{q_1+1,q_2+1}_{q} (x_{12+}, \pr_{x_{2+}}) 
\theta_{12} \x_{12+}{\tilde \pr}_{x_{2+}}\pr_{\theta_2}\cr
{}& - {q_1(q_1-1)\over 4q(q-1)} \, \theta_{12}{}^{\! 2}
C^{q_1+1,q_2}_{q}( x_{12+}, \pr_{x_{2+}})\pr_{\theta_2}{}^{\!2} \, , }
}
where ${\tilde \pr}_{x} = \tsi{\cdot \pr_x}$. The result \OPEC\ has no
singular terms for $z_{1+}\to z_{2+}$ so that the chiral superfields
form a closed algebraic ring with the scale dimension/R-charge additive.

We may use \OPEC\ to determine the contribution arising from a chiral
scalar superfield $\phi_i$ to the four point function 
$\l \phi_1(z_{1+}) \phi_2 (z_{2+}) \phi_3(z_{3+}) \phi_4(z_{4+})\r$ 
arising from the operator product expansion for
$\phi_1(z_{1+}) \phi_2 (z_{2+})$  where $q_i = q = q_1+q_2 = 3-q_3-q_4$.
{}From \tphi
\eqn\threephi{
\big \l \phi_{i}(z_{2+}) \phi_{3}(z_{3+}) \phi_{4}(z_{4+}) \big \r =
C_{i34}\,  \Lambda_{2(34)}{}^{\! 2} {
r_{34}{}^{\raise 2pt\hbox{$\scriptstyle  \! q-2$}} \over
r_{23}{}^{\raise 2pt\hbox{$\scriptstyle  \! 1-q_4$}}
r_{24}{}^{\raise 2pt\hbox{$\scriptstyle  \! 1-q_3$}}} \, .
}
and using \OPEC\ gives, after lengthy calculations extending those
which led to \OPEF\ which are described 
in appendix E, exactly the form shown in \fourp\ with
\eqn\OPER{\eqalign{
f_{123}(u,v) = {}& - C_{12{\bar i}} C_{i34} \, {q_1q_2\over q(q^2-1)}(2-q_4)
u^{q-1}G(2-q_3,q_1+1,q, q+2;u,1-v)\, , \cr
f_{124}(u,v) = {}& - C_{12{\bar i}} C_{i34} \, {q_1q_2\over q(q^2-1)}(2-q_3)
u^{q-1}G(2-q_4,q_2+1,q, q+2;u,1-v) \, ,\cr
f_{134}(u,v) = {}& C_{12{\bar \imath}} C_{i34} \, {q_1\over q}
G(q_2,2-q_4,q-1, q+1;u,1-v) \, , \cr
f_{234}(u,v) = {}& C_{12{\bar \imath}} C_{i34} \, {q_2\over q}
G(q_1,2-q_3,q-1, q+1;u,1-v) \, . }
}
This result is identical to that given by \fph, \fphb, \fphc\ and \fphd\
for a suitable choice of the coefficient $B$. These terms may therefore
be given solely by a chiral superfield in the operator product expansion
of $\phi_1,\phi_2$ and conversely the $A$ terms may be related to the
contribution of a chiral superfield in the operator product of $\phi_3,\phi_4$.

\newsec{Superconformal Integrals}

The results obtained above are illustrated by application to integrals
which automatically define superconformally covariant $N$-point functions.
Partial results were obtained by one of us earlier \HO\ and a more complete
analysis is undertaken here. As explained in the introduction, and as
exhibited in \twop, we may straightforwardly define a superconformally
covariant scalar function for two points which is chiral at $z_i$ and
anti-chiral at $z$. By integration over products of such factors for 
$i=1,\dots N$, with the appropriate weighting, and integrating over $z_-$ 
we may then obtain a chiral $N$-point function. Explicitly this gives
\eqn\IN{
I_N(z_{1+}, \dots z_{N+}) = {i\over \pi^2} \int \! \d^4 x_- \d^2 \bth \, 
\prod_{i=1}^N {\Gamma(q_i)\over (x_{i+} -2i \theta_i \si \bth - x_-
)^{\raise 2pt\hbox{$\scriptstyle 2$}}} \, ,
}
and the $q_i$ are constrained by \sumq, since this is necessary to ensure that
under a superconformal transformation the factors ${\bar \Omega}(z_-)$ from
\twop\ cancel the associated transformation of the measure in \IN. The
technique for dealing with such integrals is described in appendix B and, with
due account of analytic continuation to Minkowski space, the procedure
described there gives, for $\Lambda = \sum_i \lambda_i$,
\eqn\INa{
I_N(z_{1+}, \dots z_{N+}) =  \int_0^\infty \!\!\!\!\!
{\scriptstyle{\prod}}_{i} \d\lambda_i \lambda_i{}^{\! q_i -1}
{1\over \Lambda^2} \!
\int \!\! \d^2\bth \, \exp \Big ( -{{\textstyle{1\over \Lambda}}}
{\scriptstyle{\sum_{i<j}}} \lambda_i \lambda_j
\big (x_{i+}-x_{j+} - 2i (\theta_i - \theta_j) \si \bth\big )^2\Big ) \, .
}
The $\bth$ integration is as usual performed by expanding the exponential to
order $\bth^2$. Since $\bth$ only appears in $x_{i+}- 2i\theta_i\si \bth$
we may write
\eqn\INb{
I_N(z_{1+}, \dots z_{N+}) = -  \int_0^\infty \!\!\!\!
{\scriptstyle{\prod}}_{i} \d\lambda_i \lambda_i{}^{\! q_i -1}
{1\over \Lambda^2}
\sum_{jk} {\theta}_j \si{\cdot \pr}_j \, \tsi{\cdot \pr}_k \tth_k \,
e^{-{1\over \Lambda}\sum_{i<j} \lambda_i \lambda_j r_{ij} }  \, ,
}
for $r_{ij}$ as in \defrl. By carrying out the differentiation  \INb\ becomes
\eqn\INc{\eqalign{
I_N(z_{1+}, \dots z_{N+})= {} &  4 \int_0^\infty \!\!\!\!
{\scriptstyle{\prod}}_{i} \d\lambda_i \lambda_i{}^{\! q_i -1}
{1\over \Lambda^3}\,
e^{-{1\over \Lambda}\sum_{i<j} \lambda_i \lambda_j r_{ij}} \cr
& {}\times
\bigg( \sum_{ijk}\lambda_i\lambda_j \lambda_k\, {\theta}_i^{\vphantom g}
\, \x_{ik+}^{\vphantom g}\,  \tx_{kj+}^{\vphantom g}\, \tth_j^{\vphantom g}
- \half \sum_{ij}  \lambda_i \lambda_j \, r_{ij} \sum_k \lambda_k \,
\theta_k^{\, 2} \bigg) \, , \cr}
}
and with further rearrangement we find
\eqn\INd{
\!\! I_N(z_{1+}, \dots z_{N+})= {}  - 4 \int_0^\infty \!\!\!\!
{\scriptstyle{\prod}}_{i} \d\lambda_i \lambda_i{}^{\! q_i -1}
{1\over \Lambda^3}\, e^{-{1\over \Lambda}\sum_{i<j} \lambda_i \lambda_j r_{ij}}
\!\! \sum_{i<j<k} \!\! \lambda_i \lambda_j \lambda_k r_{ij} r_{ik} \,
\Lambda_{i(jk)}{}^{\! 2} \, .
}
The integrals appearing in \INc\ and \INd\ are of the particular form considered
by Symanzik \Symanzik\ for conformally invariant theories. His discussion is briefly
reviewed in appendix B and the essential integral, exhibited in (B.5) with (B.2),
is applicable in \INc\ and \INd\ for the particular
case of $\mu=3$. For such integrals radical simplifications are possible
so that for $N\ge 3$ they may be expressed in terms of functions of the
harmonic cross ratios of the $r_{ij}$.  For $N=3$ we have from (B.6)
\eqn\Ith{
I_3(z_{1+},z_{2+},z_{3+}) = -4\, {\Lambda_{1(23)}{}^{\! 2}\over r_{23}}\,
{\Gamma(2-q_1) \Gamma(2-q_2) \Gamma(2-q_3)
\over r_{23}{}^{\raise 2pt\hbox{$\scriptstyle \! 1 - q_1$}} \,
r_{31}{}^{\raise 2pt\hbox{$\scriptstyle \! 1 - q_2$}} \,
r_{12}{}^{\raise 2pt\hbox{$\scriptstyle \! 1 - q_3$}} } \, ,
}
which is in accord with the expected form in \tphi. For $N=4$ using (B.11)
and \sHHH\ where appropriate we have
\eqn\Ifour{\eqalign{
I_4(z_{1+},z_{2+},z_{3+},z_{4+})
={}-4\bigg \{ &{\Lambda_{1(23)}{}^{\! 2} \over r_{23}}
\bigg ( {r_{13} \over r_{23}} \bigg )^{\! q_2-1}
\bigg ( {r_{12} \over r_{23}} \bigg )^{\! q_3-1}
\bigg ( {r_{12} \over r_{24}} \bigg )^{\! q_4}  \cr
&{}\times H\big (q_4,2-q_2,q_3+q_4-1,q_3+q_4+1;u,v\big )  \cr
{}+{} &{\Lambda_{1(24)}{}^{\! 2} \over r_{24}}
\bigg ( {r_{24} \over r_{14}} \bigg )^{\! q_1-1}
\bigg ( {r_{12} \over r_{13}} \bigg )^{\! q_3}
\bigg ( {r_{12} \over r_{14}} \bigg )^{\! q_4-1}  \cr
&{}\times H\big (q_3,2-q_1,q_3+q_4-1,q_3+q_4+1;u,v\big )  \cr
{}+{} &{\Lambda_{1(34)}{}^{\! 2} \over r_{34}}
\bigg ( {r_{14} \over r_{24}} \bigg )^{\! q_2}
\bigg ( {r_{14} \over r_{34}} \bigg )^{\! q_3-1}
\bigg ( {r_{13} \over r_{34}} \bigg )^{\! q_4-1}\cr
&{}\times H\big (q_2,2-q_4,q_1+q_2-1,q_1+q_2+1;u,v\big )  \cr
{}+{} &{\Lambda_{2(34)}{}^{\! 2} \over r_{34}}
\bigg ( {r_{23} \over r_{13}} \bigg )^{\! q_1}
\bigg ( {r_{24} \over r_{34}} \bigg )^{\! q_3-1}
\bigg ( {r_{23} \over r_{34}} \bigg )^{\! q_4-1}  \cr
&{}\times H\big (q_1,2-q_3,q_1+q_2-1,q_1+q_2+1;u,v\big ) \bigg \} \, .\cr}
}
This result is identical, up to an overall constant, with the chiral four point 
function constructed in sections 3 and 4.

\newsec{Conclusion}

The results of this paper demonstrate that in some cases conformal
invariance, in conjunction with supersymmetry, is sufficient to determine
four point functions in four dimensions. Nevertheless the present results
are of limited direct relevance in that the unitarity bound for scalar
chiral superfields $q\ge 1$ is incompatible with $\sum_i q_i =3$ which
is the case when our non trivial results are obtained. Even the three
point function is applicable only if $q_i=1$ which corresponds to free
fields. However in some cases it is possible that the chiral superfields
play the role of a potential for physical fields and the unitarity bound
might not then apply. For $\N=4$ supersymmetry there are various results
which require that four and higher point functions of scalar primary
superfields are just products of free fields in some extremal cases \West. 
Perhaps for other cases as well superconformal symmetry may constrain the
functions of the conformal cross ratios $u,v$ while not requiring just
a solution corresponding to free fields.

In this context the two variable Appell functions $F_4$ together with 
the related $G$ and $H$ are a natural set of functions in terms of which 
four point correlation functions may
be expressed. The function $G$ clearly arises in the operator product
expansion and includes the contributions of all derivatives of 
quasi-primary fields. The properties  \sP, \sPP\ and \sHH, \sH, \sHHH\ are
clearly what is required to ensure the essential crossing symmetry
conditions on four point functions.

\noindent{\bigbf Acknowledgements}

One of us (FAD) would like to thank the EPSRC, the National University of
Ireland and Trinity College, Cambridge for support.

\vfill\eject
\appendix{A}{Alternative Analysis of Four Point Functions}

In terms of the discussion of section 5 we may derive an equivalent set
of equations expressing superconformal invariance. We start from
\eqn\Fourp{ \eqalign{
\big (& x_{2+}{}^{\! 2} \big )^{q_2}
\big \l \phi_1(z_{1+}) \phi_2 (z_{2+}) \phi_3 (z_{3+}) \phi_4(z_{4+}) \big \r
\Big |_{{x_{1+}=0, \theta_1 = 0 \atop x_{2+}\to \infty, \theta_2 = 0}}\cr
&{}= \big (x_{3+}{}^{\! 2}\big )^{q_2 -2} \Big (
\theta_3{}^{\! 2}\,  f_1(v,w) + \theta_4{}^{\! 2}\,  f_2(v,w) + 
2\theta_4 \tth_3 \, f_3(v,w) 
+ 2 \theta_4 \x_4\tx_3 \tth_3 {1\over x_{3+}{}^{\! 2}} \, f_4(v,w) \Big ) \, ,\cr}
}
for
\eqn\defvw{
v= {x_{4+}{}^{\! 2}\over x_{3+}{}^{\! 2}} \, , \qquad
w = {2x_{4+}{\cdot x_{3+}}\over x_{3+}{}^{\! 2}} \, .
}
In this particular limit we need consider only the variations in $z_{3+},z_{4+}$
for the terms involving $\bep,\eta$ in \sol. From terms proportional to
$i \theta_3{}^{\! 2} \theta_4 \x_4 \bep$ we have
\eqn\AA{
\pr_v f_1 - \pr_w f_3 - (v\pr_v-q_2+1) f_4 = 0 \, ,
}
and from $i \theta_3{}^{\! 2} \theta_4 \x_3 \bep$
\eqn\BB{
\pr_w f_1 + (v\pr_v + w\pr_w - q_2 + 2)f_3 - v\pr_w f_4 = 0 \, ,
}
and from  $i \theta_4{}^{\! 2} \theta_3 \x_4 \bep$
\eqn\CC{
\pr_w f_2 - \pr_v f_3 - \pr_w f_4 = 0 \, ,
}
and from $i \theta_4{}^{\! 2} \theta_3 \x_3 \bep$
\eqn\DD{
- (v\pr_v + w\pr_w - q_2 + 2) f_2 - \pr_w f_3 +
(v\pr_v + w\pr_w  + 2) f_4 = 0 \, ,
}
and from $\theta_3{}^{\! 2} \theta_4 \x_4 \tx_3 \eta$
\eqn\EE{
\pr_w f_1 + \pr_w f_3 + ( v\pr_v - q_2 - q_3 +2 ) f_4 = 0 \, ,
}
and from $\theta_4{}^{\! 2} \theta_3 \x_3 \tx_4 \eta$
\eqn\FF{
\pr_w f_2 + \pr_w f_3 - ( v\pr_v + w\pr_w + q_4 + 1 ) f_4 = 0 \, ,
}
and from $\theta_3{}^{\! 2} \theta_4 \eta$
\eqn\GG{
(v\pr_v + w \pr_w + q_4)f_1 + (v\pr_v + w \pr_w - q_2 -q_3 + 3 ) f_3 - v\pr_w f_4
= 0 \, ,
}
and from  $\theta_4{}^{\! 2} \theta_3 \eta$
\eqn\HH{
-(v\pr_v - q_2- q_3 +2)f_2 - (v\pr_v + q_4- 1 ) f_3 - v\pr_w f_4 = 0 \, .
}
For the results involving $\eta$ it is necessary to include the terms involving
$q_3 \theta_3 \eta$ and $q_4 \theta_4 \eta$ arising from $\si$ in the 
superconformal Ward identity \scf.

The coefficient functions appearing in \Fourp\ may be related to those in
\fourp\ by
\eqn\FFrel{\eqalign{
& f_{123} = f_1+f_3 - vf_4 \, , \qquad v^{q_2+q_3-2}f_{124} = f_2 + f_3 - f_4 \, , \cr
& u^{q_1+q_2-2} v^{q_2+q_3-2} f_{134} = f_4 \, , \qquad 
u^{q_1+q_2-2} f_{234} = - f_3 \, , \qquad u = 1-w+v \, . \cr}
}
With the basis in \Fourp\ the symmetry properties are less evident but the equations
obtained here are equivalent, with \FFrel, to those obtained in section 3.

\appendix{B}{Conformal Integrals}

For completeness we recapitulate some of the results of Symanzik \Symanzik\
concerning integrals defining conformally invariant $N$-point functions. For
the purposes of this appendix we assume a Euclidean metric in general
$d$ dimensions and define
\eqn\intI{
I_N(x_1,\dots x_N) = {1\over \pi^\mu} \int \! \d^d x \, 
\prod_{i=1}^N {\Gamma(\de_i) \over (x-x_i)^{2\de_i}} \, , \qquad \mu = \half d \, ,
}
and we require
\eqn\sde{
\sum_{i=1}^{N} \de_i = d = 2\mu \, ,
}
which ensures conformal invariance. Using
\eqn\GG{
{\Gamma(\de_i) \over (x-x_i)^{2\de_i}} = \int_0^\infty \!\! \d \lambda_i
\, \lambda_i{}^{\! \de_i - 1} e^{- \lambda_i (x-x_i)^2} \, ,
}
and writing $\sum_i \lambda_i (x-x_i)^2 = \Lambda \big (x-
\sum_i \lambda_i x_i /\Lambda\big )^2 + 
\sum_{i<j}\lambda_i \lambda_j r_{ij} /\Lambda$ for
\eqn\LL{
\Lambda= \sum_i \lambda_i \, , \qquad r_{ij} = (x_i - x_j)^2 \, ,
}
we may evaluate the $x$ integral to give
\eqn\Il{
I_N(x_1,\dots x_N) = \int_0^\infty \!\!\!\! {\textstyle \prod_i }
\d \lambda_i \, \lambda_i{}^{\! \de_i-1}
\, {1\over \Lambda^{\raise 2pt\hbox{$\scriptstyle \mu$}}} \, 
e^{- {1\over \Lambda} \sum_{i<j}\lambda_i \lambda_j r_{ij}} \, . 
}
The crucial observation of Symanzik is that when \sde\ holds the integral \Il\ is
unchanged if we take, instead of \LL,
 $\Lambda= \sum_i \kappa_i \lambda_i$ for any $\kappa_i \ge 0$, not all zero.
To verify this we may make the change of variables $\lambda_i =\si \alpha_i$,
with $\alpha_i$ constrained by $\sum_i \kappa_i \alpha_i = 1$. The integration
measure in \Il, $\prod_i \d \lambda_i \, \lambda^{\de_i-1} = \prod_i \d \alpha_i
\alpha_i{}^{\! \de_i-1}
\, \de(1-\sum_i \kappa_i \alpha_i) \, \d \si \, \si^{d-1}$ and then, carrying
out the integration over $\si$, the explicit dependence on $\Lambda$ disappears.
This result is then equivalent to making the above change in $\Lambda$.

For $N=3$ if we choose $\Lambda = \lambda_3$ in \Il\ then we may easily carry
out the $\lambda_3$ integration and subsequently integrate $\lambda_1,
\lambda_2$ to give \Depp
\eqn\Ithree{
I_3(x_1,x_2,x_3) = {\Gamma(\mu-\de_1) \Gamma(\mu-\de_2) \Gamma(\mu-\de_3)
\over r_{23}{}^{\raise 2pt\hbox{$\scriptstyle \! \mu - \de_1$}} \, 
r_{31}{}^{\raise 2pt\hbox{$\scriptstyle \! \mu - \de_2$}} \,
r_{12}{}^{\raise 2pt\hbox{$\scriptstyle \! \mu - \de_3$}} } \, .
}

For $N=4$ we choose $\Lambda = \lambda_4$ and, following Symanzik \Symanzik,
write
\eqn\st{\eqalign{
e^{-{1\over \lambda_4}\lambda_1 \lambda_2 r_{12}} ={}&
{1\over 2\pi i} \int_{c-i\infty}^{c+i\infty}\!\!\!\!\!\!\!\!\!\! \d s \, \, 
\Gamma(-s) \bigg ( {\lambda_1 \lambda_2 \over \lambda_4}r_{12} \bigg )^{\! s} \, ,
\cr e^{-{1\over \lambda_4}\lambda_2 \lambda_3 r_{23}} ={}&
{1\over 2\pi i} \int_{c-i\infty}^{c+i\infty}\!\!\!\!\!\!\!\!\!\! \d t \, \, 
\Gamma(-t) \bigg ( {\lambda_2 \lambda_3 \over \lambda_4}r_{23}\bigg )^{\! t}\, ,
\qquad c<0 \, , \cr}
}
and then we can evaluate the $\lambda_4$ and subsequently the $\lambda_1,
\lambda_2,\lambda_3$  integrations to obtain
\eqn\IF{
I_4(x_1,x_2,x_3,x_4) = r_{14}{}^{\raise 2pt\hbox{$\scriptstyle \! \mu - \de_1
- \de_4$}}\,  r_{34}{}^{\raise 2pt\hbox{$\scriptstyle \! \mu - \de_3 - \de_4$}}
r_{13}{}^{\raise 2pt\hbox{$\scriptstyle \! - \mu + \de_4$}} \, 
r_{24}{}^{\raise 2pt\hbox{$\scriptstyle \! - \de_2$}}\, \F_4(u,v) \, ,
}
where $u,v$ are as in \defuv\ and
\eqn\Fint{\eqalign{
\F_4(u,v) = {1\over (2\pi i)^2} \int \!\! \d s \, \d t \, &
\Gamma(-s) \Gamma(-t) \Gamma(s+t+\de_2) \Gamma(s+t+\mu - \de_4)\cr
&{}\times \Gamma(\de_3+\de_4-\mu-s) \Gamma(\de_1+\de_4-\mu -t) \, u^s v^t \, .\cr}
} 
By closing the contours in ${\sl Re}\, s, {\sl Re}\, t > 0$, using
$\Gamma(x-m)= \Gamma(x)(-1)^m/(1-x)_m$, the integrals
may be obtained in terms of the $F_4$ functions defined by \defF\foot{Such
functions were introduced also in a related context by Ferrara {\it et al}
in \Ferr\ and the essential integral \Fint\ was considered by
Davydychev and Tausk \Dav\ who also found its expression in terms
of $F_4$ functions.}
\eqn\FFfour{\eqalign{ \!\!\!\!\!
\F_4(u,v) & =  \Gamma(\de_2) \Gamma(\mu-\de_4)\Gamma(\de_3+\de_4-\mu)
\Gamma(\de_1+\de_4-\mu) \cr
&\ {}\times 
F_4\big (\de_2,\mu-\de_4,\mu+1-\de_3-\de_4,\mu+1-\de_1-\de_4;u,v\big ) \cr
&\ {} + \Gamma(\mu-\de_3) \Gamma(\de_1)\Gamma(\de_3+\de_4-\mu)
\Gamma(\de_2+\de_3-\mu) \cr
&\ {} \times v^{\de_1+\de_4-\mu}
F_4\big (\mu-\de_3,\de_1,\mu+1-\de_3-\de_4,\mu+1-\de_2-\de_3;u,v\big ) \cr
&\ {} + \Gamma(\mu-\de_1) \Gamma(\de_3)\Gamma(\de_1+\de_2-\mu)
\Gamma(\de_1+\de_4-\mu) \cr
&\ {} \times u^{\de_3+\de_4-\mu}
F_4\big (\mu-\de_1,\de_3,\mu+1-\de_1-\de_2,\mu+1-\de_1-\de_4;u,v\big ) \cr
&\ {} + \Gamma(\de_4) \Gamma(\mu-\de_2)\Gamma(\de_1+\de_2-\mu)
\Gamma(\de_2+\de_3-\mu) \cr
&\ {} \times u^{\de_3+\de_4-\mu}v^{\de_1+\de_4-\mu}
F_4\big (\de_4,\mu-\de_2,\mu+1-\de_1-\de_2,\mu+1-\de_2-\de_3;u,v\big ) \, . \cr }
}
With the aid of \defG\ and \defH\ this may be reduced just to the function $H$
so that
\eqn\IH{ \eqalign{
I_4(x_1,x_2,x_3,x_4) ={}&{1\over r_{13}{}^{\raise 2pt\hbox{$\scriptstyle \! \mu$}}}
\bigg ( {r_{14} \over r_{24}} \bigg )^{\! \de_2}
\bigg ( {r_{14} \over r_{34}} \bigg )^{\! \de_3-\mu}
\bigg ( {r_{13} \over r_{34}} \bigg )^{\! \de_4}  \cr
&{}\times
H\big (\de_2,\mu-\de_4,\de_1+\de_2+1-\mu,\de_1+\de_2;u,v\big ) \cr
={}&{1\over r_{13}{}^{\raise 2pt\hbox{$\scriptstyle \! \mu$}}}
\bigg ( {r_{13} \over r_{23}} \bigg )^{\! \de_2}
\bigg ( {r_{12} \over r_{23}} \bigg )^{\! \de_3-\mu}
\bigg ( {r_{12} \over r_{24}} \bigg )^{\! \de_4}  \cr
&{}\times
H\big (\de_4,\mu-\de_2,\de_3+\de_4+1-\mu,\de_3+\de_4;u,v\big ) \, , \cr}
}
where the two expressions are related by \sHHH.\foot{By considering the limit
$x_4{}^{\! 2}\to \infty$, when $u= r_{12}/r_{13}, \, v=r_{23}/r_{13}$, this
also gives an expression for the integral $I_3$ in \intI\ without the
condition \sde\ which then determines $\de_4$ in \IH.}

\appendix{C}{Particular Results for $G$ and $H$}

The two variable functions $F_4$, and also $G$, $H$ given by \defG, \defH, are
relatively unfamiliar although $H$ occurs in various contexts in the
evaluation of Feynman integrals \refs{\Dav,\Uss}. In this appendix, we list
some results for $G$ which are relevant and give formulae for special
cases, when they reduce to well known single variable functions.

{}From the definition \defG\ we may directly obtain
\eqn\dG{\eqalign{
\pr_u G(\alpha,\beta,\gamma,\delta;u,1-v) = {}& {\alpha\beta(\de-\alpha)
(\de-\beta)\over \gamma \de(\de+1)} \cr
{}& \times G(\alpha+1,\beta+1,\gamma+1,\delta+2;u,1-v)  \, , \cr
\pr_v G(\alpha,\beta,\gamma,\delta;u,1-v) = {}& - {\alpha\beta\over\de}\,
G(\alpha+1,\beta+1,\gamma,\delta+1;u,1-v) \, , \cr
\big (\beta+u\pr_u + v\pr_v \big ) G(\alpha,\beta,\gamma,\delta;u,1-v)
= {}& {\beta(\de-\alpha)\over \de} \,
G(\alpha,\beta+1,\gamma,\delta+1;u,1-v) \, . }
}
There are also various recurrence relations:
\eqna\rG$$
\eqalignno{\!\!\!\!\!\!
\de (\de-\gamma - \beta+1) G(\alpha,\beta,\gamma,\delta;u,1-v) = {}&
- \de (\gamma-1)G(\alpha,\beta,\gamma-1,\delta;u,1-v) \cr
{}& + \alpha(\de-\beta)
G(\alpha+1,\beta,\gamma,\delta+1;u,1-v) \cr
{}& + (\de-\alpha)(\de-\beta) G(\alpha,\beta,\gamma,\delta+1;u,1-v) \, , &
\rG a\cr
\de G(\alpha,\beta,\gamma,\delta;u,1-v) =  \beta
G(\alpha,\beta+1,\gamma,& \delta+1;u,1-v) +(\de-\beta)
G(\alpha,\beta,\gamma,\delta+1;u,1-v) \cr
- {\alpha\beta(\de-\beta)\over \gamma(\de+1)}
\, u & G(\alpha+1,\beta+1,\gamma+1,\delta+2;u,1-v) \, , & \rG b \cr
(\beta-\alpha)\de G(\alpha,\beta,\gamma,\delta;u,1-v) = {}&
(\de-\alpha)\beta G(\alpha,\beta+1,\gamma,\delta+1;u,1-v) \cr
{}& - (\de-\beta)\alpha G(\alpha+1,\beta,\gamma,\delta+1;u,1-v) \, ,& \rG c \cr
\de (\de-\alpha - \beta) G(\alpha,\beta,\gamma,\delta;u,1-v) = {}&
(\de - \alpha) (\de-\beta)G(\alpha,\beta,\gamma,\delta+1;u,1-v) \cr
{}& - \alpha\beta \, vG(\alpha+1,\beta+1,\gamma,\delta+1;u,1-v) \, . & \rG d }
$$

For the particular case when $\de= \gamma+n$, $n=0,1,\dots$ the $G$-function
may be reduced to products of ordinary hypergeometric functions. 
Defining\foot{To invert we may define $\lambda =
( (1-u-v)^2 - 4uv )^{1\over 2}=1-x-y$ and 
$\rho=2/(1-u-v +\lambda) = 1/((1-x)(1-y))$ and then
$x=\rho u /(1+\rho u)$, $y=\rho v /(1+\rho v)$.}
\eqn\uvxy{
u = x(1-y) \, , \qquad v=y(1-x) \, ,
}
these originate from the reduction formula for  $F_4$
\eqn\red{
F_4(\alpha,\beta,\gamma,\gamma';u,v)= F(\alpha,\beta;\gamma;x)
F(\alpha,\beta;\gamma';y) \, , \quad \alpha+\beta = \gamma+\gamma'-1 \, .
}
Applying this in \PF\ and using standard hypergeometric identities gives
\eqn\redG{
G(\alpha,\beta,\gamma,\gamma;u,1-v)= F(\alpha,\beta;\gamma;x)
F(\alpha,\beta;\gamma;1-y) \, .
}
Using \dG\ for $\pr_v G$ then gives
\eqn\redGG{\eqalign{\!\!\!\!\!
G(\alpha,\beta,\gamma & ,\gamma+1;u,1-v)\cr
= {}&  {1\over 1-x-y} \big (
(1-y) F(\alpha-1,\beta-1;\gamma;x)F(\alpha,\beta;\gamma+1;1-y) \cr
{}& \qquad \qquad\quad
- x F(\alpha,\beta;\gamma+1;x)F(\alpha-1,\beta-1;\gamma;1-y) \big ) \cr
= {}& {1\over 1-x-y}{\gamma\over \alpha-1} \big ( F(\alpha-1,\beta-1;\gamma;x)
F(\alpha-1,\beta;\gamma;1-y) \cr
{}& \qquad \qquad\qquad\quad - F(\alpha,\beta-1;\gamma;x)
F(\alpha-1,\beta-1;\gamma;1-y) \big ) \cr
= {}& {1\over 1-x-y}{\gamma(\gamma-1)\over (\alpha-1)(\beta-1)} 
\big ( F(\alpha-1,\beta-1;\gamma;x)
F(\alpha-1,\beta-1;\gamma-1;1-y) \cr
{}& \qquad \qquad\qquad\qquad\quad  \ - F(\alpha-1,\beta-1;\gamma-1;x)
F(\alpha-1,\beta-1;\gamma;1-y) \big ) \, .}
}
The results for the four point functions obtained here are given by the
$G$-function for $\de=\gamma+2$. Following a similar route as which led to
\redGG\ we may obtain
\eqn\redGGG{\eqalign{
G& (\alpha+1,\beta+1,\gamma ,\gamma+2;u,1-v)\cr
{}& = {1\over (1-x-y)^2} \Big ( (1-y)^2 
F(\alpha-1,\beta-1;\gamma;x)F(\alpha+1,\beta+1;\gamma+2;1-y)\cr
{}& \qquad \qquad \qquad\qquad  + x^2
F(\alpha+1,\beta+1;\gamma+2;x) F(\alpha-1,\beta-1;\gamma;1-y) \Big ) \cr
{}& - 2{\gamma+1 \over \alpha\beta} \, {x(1-y)\over (1-x-y)^3}\cr
{}& \ \times \Big (
\big ( \gamma F(\alpha{-1},\beta{-1};\gamma;x) - (\gamma-1)
F(\alpha{-1},\beta{-1};\gamma{-1};x) \big ) F(\alpha,\beta;\gamma{+1};1-y) \cr
{}& \quad - F(\alpha,\beta;\gamma{+1};x)
\big ( \gamma F(\alpha{-1},\beta{-1};\gamma; 1-y)
- (\gamma-1) F(\alpha{-1},\beta{-1};\gamma{-1};1-y) \big )\Big )  \, . }
}

The result \redG\ is relevant, according to \FFfour, in two dimensions.
Using complex coordinates for this case and defining
\eqn\deta{
\eta = {(z_1-z_2)(z_3-z_4)\over (z_1-z_3)(z_2-z_4)} \, ,
}
we have
\eqn\uveta{
u = \eta {\bar \eta} \, , \qquad v = (1-\eta)(1-{\bar \eta})\, ,
}
and then from \redG\ we have
\eqn\Gtwo{
G(\alpha,\beta,\gamma,\gamma;u,1-v) = F(\alpha,\beta;\gamma;\eta)
F(\alpha,\beta;\gamma; {\bar \eta} ) \, ,
}
exhibiting a holomorphic factorisation. For the crossing symmetric
function $H$ given by \defH\ we have
\eqn\Htwo{\eqalign{
H(\alpha,\beta,\gamma,\gamma;u,v) = {\pi \over \sin \pi \gamma}
\Big \{& N_1 \big | F(\alpha,\beta;\gamma;\eta) \big |^2 \cr
{}& + N_2 \big | \eta^{1-\gamma}
F(\alpha-\gamma+1,\beta-\gamma+1;2-\gamma;\eta) \big |^2 \Big \} \, , \cr
N_1 = {\Gamma(\alpha)\Gamma(\beta)\Gamma(\gamma-\alpha)
\Gamma(\gamma-\beta)\over \Gamma(\gamma)^2}& \, , \quad
N_2 = -{\Gamma(1-\alpha)\Gamma(1-\beta)\Gamma(\alpha-\gamma+1)
\Gamma(\beta-\gamma+1)\over \Gamma(2-\gamma)^2} \, ,
\cr}
}
which coincides with expressions previously obtained for 
two dimensional conformal four point functions \Dot. 

We may also apply \redGG\ to cases of relevance in four dimensions. 
It is easy to see that it gives
\eqn\Gfour{
G(1,1,1,2;u,1-v) =  - {1\over 1-x-y}
\ln {y\over 1-x} \, .
}
The definition \defH\ for $H$ gives
\eqn\HG{\eqalign{
H(1,1,1+\epsilon,2+\epsilon;u,v) = {\pi \over \sin \pi \epsilon} \bigg (&
- {1\over 1+\epsilon} G(1,1,1+\epsilon,2+\epsilon;u,1-v)\cr
& + {1\over 1-\epsilon} \Big ( {v\over u} \Big )^{\! \epsilon}
G(1,1,1-\epsilon,2-\epsilon;u,1-v) \bigg ) \, , }
}
and using \redGG\ with
\eqn\expF{
wF(1,1;2+\ep;w) = -(1+\ep) \ln (1-w) - \ep
\Big ( \half \big ( \ln(1-w)\big)^2 + {\rm Li}_2(w) \Big ) + {\rm O}
(\ep^2) \, ,
}
where ${\rm Li}_2$ is the dilogarithm function, we have
\eqn\Hfour{
H(1,1,1,2;u,v) =  {1\over 1-x-y} \Big ( \ln x(1-y) \, \ln {y\over 1-x} 
- 2{\rm Li}_2 (x) + 2{\rm Li}_2(1-y) \Big ) \, .
}
This result is  relevant acording to \IH\ if  $N,d=4$ in \intI\ and we assume
$\de_1=\de_2=\de_3=\de_4=1$ and is equivalent with the the results given in
\refs{\Dav,\Uss} for this integral.

It is of interest to verify that \Hfour\ satisfies the symmetry relations
\sHH\ and \sH. For $u \leftrightarrow v$ when $x \leftrightarrow y$
we may use  ${\rm Li}_2 (x) + {\rm Li}_2 (1-x) = {1\over 6}\pi^2 - \ln x \,
\ln(1-x)$ whereas when $u\to u' = u/v, \, v\to v'=1/v$ we have
$x'=1- 1/y, \, y'=1/(1-x)$ we may use the identity
${\rm Li}_2 (x) + {\rm Li}_2(x/(x-1)) = - \half (\ln(1-x))^2$. 

\appendix{D}{Operator Product Expansion Calculations}

We describe here some details of the calculations involved in applying
the operator product expansion to the four point function. The essential
derivative operator defined by \defC\ is given explicitly by
\eqn\expC{\eqalign{
C^{a,b}_{\kappa}(s,\pr) = {}& {1\over B(a,b)}\sum_{n=0}{1\over n!(\kappa)_n}
\big ( {- \quar s^2 \pr^2} \big )^n \int_0^1 \!\!\! \d \alpha \,
\alpha^{a+n-1}(1-\alpha)^{b+n-1} e^{\alpha s{\cdot \pr}} \cr
= {}& {1\over B(a,b)}\sum_{m,n=0} {(a)_{m+n} (b)_{n} \over m!n!
(a+b)_{m+2n} (\kappa)_n} \, \big ( s{\cdot \pr} \big )^m 
\big ( {- \quar s^2 \pr^2} \big )^n \, . }
}
To verify \defC\ we may use
\eqn\prx{
\big ( \quar \pr_{x_2}{}^{\! 2} \big )^n 
{1\over r_{23}{}^{\raise 2pt\hbox{$\scriptstyle  \! S$}}} ={}
(S)_n (S+1-\half d)_n
{1\over r_{23}{}^{\raise 2pt\hbox{$\scriptstyle  \! S+n$}}}\, ,
}
so that
\eqn\Ccal{
C^{a,b}_{S+1 - {1\over 2}d}(x_{12},\pr_{x_2})
{1\over r_{23}{}^{\raise 2pt\hbox{$\scriptstyle  \! S$}}} =
{1\over B(a,b)}\int_0^1 \!\!\! \d \alpha \, \alpha^{a-1}(1-\alpha)^{b-1} 
\sum_{n=0} {(S)_n \over n! } {\big ( {- \alpha(1-\alpha) r_{12}}
\big )^n \over \big ( x_{23} + \alpha x_{12} \big )^{2(S+n)}} \, .
}
With $( x_{23} + \alpha x_{12})^2 = (1-\alpha) r_{23} + \alpha r_{13} 
- \alpha(1-\alpha)r_{12}$
the sum over $n$ is straightforward, giving $\big ( (1-\alpha) r_{23}
+ \alpha r_{13} \big )^{-S}$,  and then the $\alpha$ integration, with
$S=a+b$, is of the form
\eqn\inta{
\int_0^1 \!\!\! \d \alpha \, \alpha^{a-1}(1-\alpha)^{b-1}
{1\over \big ( \alpha s + (1-\alpha) t\big )^{a+b}} = 
B(a,b) \, {1\over s^a t^b} \, ,
}
giving the desired result.

To sketch the calculation leading to \OPEF\ we follow the method described in
\Pet\ (although related results were obtained long ago in \Fone) and first 
write, by applying similar methods as which led to \Ccal,
\eqn\Cef{\eqalign{ \!\!\!\!\!
C^{a,b}_{S+1 - {1\over 2}d}& (x_{12},\pr_{x_2})
{1\over r_{23}{}^{\raise 2pt\hbox{$\scriptstyle \! e$}}
r_{24}{}^{\raise 2pt\hbox{$\scriptstyle \! f$}}} \cr
={}& {1\over B(a,b)B(e,f)} \int_0^1 \!\!\! \d \alpha \,
\alpha^{a-1}(1-\alpha)^{b-1} \int_0^1 \!\!\! \d \beta \,
\beta^{e-1}(1-\beta)^{f-1} \cr
{}& \times \!\!
\sum_{m,n=0} {(S)_{m+n} (S+1-\half d)_{m+n}\over m! n! (S+1-\half d)_m
(S+1-\half d)_n} {A^m B^n \over 
\big (\alpha x_{12} - \beta x_{34} + x_{24}\big )^{2(S+m+n)}} \, , \cr
& \qquad A= - \alpha(1-\alpha)r_{12}\, , \qquad \qquad
B= - \beta(1-\beta)r_{34} \, .}
}
By using $(\alpha x_{12} - \beta x_{34} + x_{24})^2= A+B+C$ where
$C=\alpha(1-\beta)r_{14} + \beta(1-\alpha)r_{23} + \alpha\beta r_{13}
+ (1-\alpha)(1-\beta)r_{24}$ we may express the second line on the r.h.s.
of \Cef\ in the form
\eqn\CR{
{1\over C^S}\sum_{m=0} {(S)_{2m}  \over m! (S+1-\half d)_m}
\, \Big ( {AB\over C^2} \Big )^{\! m} \, .
}
The $\alpha$ integration is just as in \inta\ and the $\beta$ integration 
may be carried out using
\eqn\intb{
\!\!\!\! \int_0^1 \!\!\! \d \beta \,\beta^{e-1}(1-\beta)^{f-1} 
{1\over (1-\beta x)^a (1-\beta y)^b} = 
B(e,f) {1\over (1-x)^{e}} \sum_{n=0} {(e)_n (b)_n \over n! (S)_n} 
\Big ( {y-x\over 1-x} \Big )^{\! n} \, , 
}
where $S=a+b=e+f$ and we apply this result for $x=1-r_{13}/r_{14}, \,
y=1-r_{23}/r_{24}$ so that $(y-x)/(1-x)=1-v$. Combining these expressions
gives  \OPEF\ with $G$ given by the series in \defG.

These results simplify in two dimensions since, using complex coordinates,
\expC\ factorises,
\eqn\twoC{
C^{a,b}_{S}(s,\pr) = {}_1\! F_1(a;S;s_z \pr_z)
\,{}_1\! F_1(a;S;s_{\bar z} \pr_{\bar z} ) \, , \qquad S=a+b \, ,
}
since
\eqn\Fone{
{}_1\! F_1(a;S;z_{12} \pr_{z_2}) {1\over z_{23}^{\ \,S}} = 
{1\over z_{12}^{\ \,a} \, z_{23}^{\  \,b}} \, .
}
For this case
\eqn\Ftwo{
{}_1\! F_1(a;S;z_{12} \pr_{z_2}) {1\over z_{23}^{\ \,e} \, z_{24}^{\ \,f}}
= {1\over z_{14}^{\ \,a}\, z_{24}^{\ \,b}} 
\Big ( {z_{14}\over z_{13}} \Big )^{\! e} F(b,e;S;\eta) \, ,
}
with $\eta$ defined by \deta. This result is in accord with the holomorphic
factorisation required by \redG.

For application to the supersymmetric case we also consider, when $d=4$,
the extension to spinor fields. For this we require
\eqn\defCS{
C^{a,b}(x_{12},\pr_{x_2})_\alpha{}^\beta\,
{(\x_{23})_{\beta\dal}\over r_{23}{}^{\raise 2pt\hbox{$\scriptstyle \! S+1$}}} 
= {(\x_{13})_{\alpha\dal}\over r_{13}{}^{\raise 2pt\hbox{$\scriptstyle \! a+1$}}
r_{23}{}^{\raise 2pt\hbox{$\scriptstyle \! b$}}} \, .
}
This has the solution
\eqn\solC{
C^{a,b}(x_{12},\pr_{x_2})_\alpha{}^\beta = C^{a+1,b}_{S-1}(x_{12},\pr_{x_2})
\de_\alpha{}^\beta +{b\over S^2-1} C^{a+1,b+1}_{S}(x_{12},\pr_{x_2}) 
\half  ( \x_{12}{\tilde \pr}_{x_{2}} )_\alpha{}^\beta \, .
}
and, in a similar fashion to the derivation of \OPEF,
\eqn\OPEG{\eqalign{
C^{a,b}(x_{12},\pr_{x_2})_\alpha{}^\beta &\,
{(\x_{23})_{\beta\dal}\over r_{23}{}^{\raise 2pt\hbox{$\scriptstyle \! e+1$}}
r_{24}{}^{\raise 2pt\hbox{$\scriptstyle \! f$}}}\cr
={}& {1\over r_{14}{}^{\raise 2pt\hbox{$\scriptstyle \! a+1$}}
r_{24}{}^{\raise 2pt\hbox{$\scriptstyle \! b$}}} \bigg ({r_{14}\over r_{13}}
\bigg )^{\! e+1} \bigg \{ (\x_{13})_{\alpha\dal} G(b,e+1,S-1,S+1; u,1-v) \cr
{}&\ \quad + {bf\over S^2-1} \, (\x_{12}\x_{42}{}^{\!-1} \x_{43})_{\alpha\dal}
G(b+1,e+1,S,S+2; u,1-v) \bigg \} \, . }
}

\appendix{E}{Supersymmetric Calculations}

The application of the operator product expansion to the supersymmetric case
requires an extension of the previous results. We first derive the
expression \CCC\ which is required to satisfy \OPES. To achieve this we
use
\eqn\expand{
{1 \over\big (x - 2i \theta_{12} \si \bth )
{}^{\raise 2pt\hbox{$\scriptstyle  2q_1$}}} =
{1 \over x{}^{\raise 2pt\hbox{$\scriptstyle  2q_1$}}} + 4iq_1 \,
{\theta_{12}\x \bth \over x{}^{\raise 2pt\hbox{$\scriptstyle  2(q_1+1)$}}}
+ 4q_1(q_1-1) \, {\theta_{12}{}^{\! 2} \bth^2 
\over x{}^{\raise 2pt\hbox{$\scriptstyle 2(q_1+1)$}}} \, ,
}
in \OPES\ with $x=x_{1+} - 2i \theta_2 \si \bth - x_-$. The results \defC\
and \defCS\ then show, since $-\quar \pr_{\theta_2}{}^{\!2} \theta_2{}^{\!2}
=1$,  that \OPES\ is satisfied if
\eqn\CC{
\eqalign{
\C^{q_1,q_2}(z_{12+}, \pr_{z_{2+}}) ={}& C^{q_1,q_2}_{q-1}
(x_{12+}, \pr_{x_{2+}})
+ {q_1\over q} \, \theta_{12} C^{q_1,q_2} (x_{12+}, \pr_{x_{2+}}) 
\pr_{\theta_2}\cr
{}& - {q_1(q_1-1)\over 4q(q-1)} \theta_{12}{}^{\! 2}
C^{q_1+1,q_2}_{q}( x_{12+}, \pr_{x_{2+}})\pr_{\theta_2}{}^{\!2} \, , }
}
which is identical with \CCC.

The results in \OPER\ are obtained by calculating
\eqn\Sfour{
\C^{q_1,q_2}(z_{12+}, \pr_{z_{2+}}) {\Lambda_{2(34)}{}^{\! 2}\over  
r_{23}{}^{\raise 2pt\hbox{$\scriptstyle  \! 1-q_4$}}
r_{24}{}^{\raise 2pt\hbox{$\scriptstyle  \! 1-q_3$}}} \, 
r_{34}{}^{\raise 2pt\hbox{$\scriptstyle  \! q-2$}} \, .
}
The terms ${\rm O}(\theta_{12}{}^{\! 2})$ are easily seen, using \OPEF, to be
\eqn\Cone{\eqalign{
& {q_1(q_1-1)\over q(q-1)}\, \theta_{12}{}^{\! 2} \,
C^{q_1+1,q_2}_{q}( x_{12+}, \pr_{x_{2+}}) 
{r_{34}{}^{\raise 2pt\hbox{$\scriptstyle  \! q-1$}} \over
r_{23}{}^{\raise 2pt\hbox{$\scriptstyle  \! 1-q_4$}}
r_{24}{}^{\raise 2pt\hbox{$\scriptstyle  \! 1-q_3$}}} \cr
{}& = {q_1(q_1-1)\over q(q-1)}\, \theta_{12}{}^{\! 2}
{r_{34}{}^{\raise 2pt\hbox{$\scriptstyle  \! q-1$}} \over
r_{14}{}^{\raise 2pt\hbox{$\scriptstyle  \! q_1+1$}}
r_{24}{}^{\raise 2pt\hbox{$\scriptstyle  \! q_2$}}} 
\Big ( {r_{14}\over r_{13}} \Big )^{\! 2-q_4} \!
G(2-q_4,q_2,q,q+1;u,1-v) \, , }
}
where, using \sP\ and \rG{a},
\eqnn\Got
$$\eqalignno{
{1-q_1\over q_2} G(2-q_4,q_2,q,q+1& ;u,1-v) = 
{1-q_1\over q_2} v^{q_1+q_4-1} G(2-q_3,q_1+1,q,q+1;u,1-v)\cr
={}& {2-q_4\over q+1} \, v^{q_1+q_4-1}G(2-q_3,q_1+1,q, q+2;u,1-v) \cr
{}& +{2-q_3\over q+1}\,  G(2-q_4,q_2+1,q, q+2;u,1-v)\cr
{}& + {1\over q_2}(1-q) \, G(2-q_4,q_2,q-1, q+1;u,1-v) \, . &\Got}
$$

The ${\rm O}(\theta_{12})$ terms arise from
\eqn\theone{
2{q_1\over q} \, \theta_{12}C^{q_1,q_2}(x_{12+},\pr_{x_{2+}})
{\tth_{23} + \x_{23+} {\tilde \ell}_{34} \over 
r_{23}{}^{\raise 2pt\hbox{$\scriptstyle  \! 2-q_4$}}
r_{24}{}^{\raise 2pt\hbox{$\scriptstyle  \! 2-q_3$}}}\,
r_{34}{}^{\raise 2pt\hbox{$\scriptstyle  \! q-1$}}\, ,
}
with $C^{q_1,q_2}(x_{12+},\pr_{x_{2+}})$ given by \solC. For the terms
not involving $\theta_{23}$ \OPEG\ gives
\eqn\thetwo{\eqalign{
{}& 2{q_1\over q} \, {r_{34}{}^{\raise 2pt\hbox{$\scriptstyle  \! q-1$}}
\over r_{14}{}^{\raise 2pt\hbox{$\scriptstyle \! q_1+1$}}
r_{24}{}^{\raise 2pt\hbox{$\scriptstyle \! q_2$}}} \bigg ({r_{14}\over r_{13}}
\bigg )^{\! 2-q_4} \bigg \{ \theta_{12}\x_{13+} {\tilde \ell}_{34} \,
G(q_2,2-q_4,q-1,q+1; u,1-v) \cr
{}& \qquad \ + {q_2(2-q_3)\over q^2-1} \, 
\theta_{12}\x_{12+}\x_{42+}{}^{\!\!\!-1} \x_{43+} {\tilde \ell}_{34} \, 
G(q_2+1,2-q_4,q,q+2; u,1-v) \bigg \} \, . }
}
The remaining terms involving $\theta_{23}$ may be calculated from this
using
\eqn\xde{
{(\x_{23+})_{\beta\dbe} \over 
r_{23}{}^{\raise 2pt\hbox{$\scriptstyle \! 2-q_4$}}} 
\, \big (\tsi{\cdot{\overleftarrow {\pr}}_{\!\!x_{3+}}}\big ){}^{\dbe\alpha} 
= 2q_4 \, {1\over r_{23}{}^{\raise 2pt\hbox{$\scriptstyle \! 2-q_4$}}} \, 
\de_\beta{}^{\! \alpha} \, .
}
Using both the derivative relations \dG\ and \rG{b} we get
\eqn\thethree{\eqalign{
{}& 2{q_1\over q} \, {r_{34}{}^{\raise 2pt\hbox{$\scriptstyle  \! q-1$}}
\over r_{14}{}^{\raise 2pt\hbox{$\scriptstyle \! q_1+1$}}
r_{24}{}^{\raise 2pt\hbox{$\scriptstyle \! q_2$}}} \bigg ({r_{14}\over r_{13}}
\bigg )^{\! 2-q_4} \bigg \{ \theta_{12}\tth_{23} \,
G(q_2,2-q_4,q-1,q+1; u,1-v) \cr
{}& \qquad \ \ + {q_2(2-q_4)\over q^2-1} \,
\theta_{12}\x_{12+}\x_{32+}{}^{\!\!\!-1} \tth_{23} \,
v^{q_1+q_4-1} G(q_1+1,2-q_3,q,q+2; u,1-v) \cr
{}& \qquad \ \ + {q_2(2-q_3)\over q^2-1} \,
\theta_{12}\x_{12+}\x_{42+}{}^{\!\!\!-1} \tth_{23} \,
G(q_2+1,2-q_4,q,q+2; u,1-v) \bigg \}  \, . }
}

The remaining terms arise from
\eqn\thefour{
C^{q_1,q_2}_{q-1}( x_{12+}, \pr_{x_{2+}})
{\Lambda_{2(34)}{}^{\! 2} \over
r_{23}{}^{\raise 2pt\hbox{$\scriptstyle  \! 1-q_4$}}
r_{24}{}^{\raise 2pt\hbox{$\scriptstyle  \! 1-q_3$}}}\, 
r_{34}{}^{\raise 2pt\hbox{$\scriptstyle  \! q-2$}} \, ,
}
where it is convenient to write
\eqn\sLam{
\Lambda_{2(34)}{}^{\! 2} =
{1\over r_{23}r_{24}} \big (\theta_{23}{}^{\! 2} r_{24} + 
( \theta_{24}{}^{\! 2}- 2 \theta_{23}\tth_{24}) r_{23} 
+ 2  \theta_{23} \x_{23+} \tx_{34+} \tth_{24} \big ) \, .
}
The $\theta_{23}{}^{\! 2}$ and $\theta_{24}{}^{\! 2}$ terms are found
from \OPEF\ to be
\eqn\thefive{
{r_{34}{}^{\raise 2pt\hbox{$\scriptstyle  \! q-2$}} \over
r_{14}{}^{\raise 2pt\hbox{$\scriptstyle  \! q_1$}}
r_{24}{}^{\raise 2pt\hbox{$\scriptstyle  \! q_2$}}}
\Big ( {r_{14}\over r_{13}} \Big )^{\! 1-q_4}\! \bigg \{ {r_{14}\over r_{13}}
\theta_{23}{}^{\! 2} G(2-q_4,q_2,q-1,q;u,1-v) 
+ \theta_{24}{}^{\! 2} G(1-q_4,q_2,q-1,q;u,1-v) \bigg \} \, . 
}
Using \sP\ and \rG{b} we may rewrite the $G$-functions as
\eqn\Ggt{\eqalign{
G(2-q_4,q_2,& q-1,q;u,1-v) = 
{q_1\over q}  G(2-q_4,q_2,q-1,q+1;u,1-v)\cr
{}& + {q_2\over q} v^{q_1+q_4-2}G(2-q_3,q_1,q-1,q+1;u,1-v) \cr
{}& -{q_1q_2(2-q_4)\over q(q^2-1)}\, u  v^{q_1+q_4-2}
G(2-q_3,q_1+1,q, q+2;u,1-v) \, , \cr
G(1-q_4,q_2,& q-1,q;u,1-v) = {q_1\over q}  G(2-q_4,q_2,q-1,q+1;u,1-v)\cr
{}& + {q_2\over q} v^{q_1+q_4-1}G(2-q_3,q_1,q-1,q+1;u,1-v) \cr
{}& -{q_1q_2(2-q_3)\over q(q^2-1)}\, u 
G(2-q_4,q_2+1,q,q+2;u,1-v) \, . }
}
For the other terms present in \sLam\ we may use
\eqn\rest{
{\x_{23} \over r_{23}{}^{\raise 2pt\hbox{$\scriptstyle  \! 2-q_4$}}}
= {1\over 2(1-q_4)} \, {1\over
r_{23}{}^{\raise 2pt\hbox{$\scriptstyle  \! 1-q_4$}}} 
\sigma {\cdot {\overleftarrow \pr}_{ \! 3}} \, .
}
{}From the derivative of the terms $\propto \theta_{24}{}^{\! 2}$ in
\thefive\ we then obtain
\eqn\thesix{\eqalign{\!\!\!\!\!\!
2{r_{34}{}^{\raise 2pt\hbox{$\scriptstyle  \! q-2$}} \over
r_{14}{}^{\raise 2pt\hbox{$\scriptstyle  \! q_1$}}
r_{24}{}^{\raise 2pt\hbox{$\scriptstyle  \! q_2$}}}
\Big ( {r_{14}\over r_{13}} \Big )^{\! 1-q_4}& \! \bigg \{ 
\theta_{23}\x_{13+}\x_{41+}{}^{\!\!\!-1} \tth_{24} \, {q_1\over q} 
G(2-q_4,q_2,q-1,q+1;u,1-v)\cr
{}& + \theta_{23}\x_{23+}\x_{42+}{}^{\!\!\!-1} \tth_{24} \, {q_2\over q}
G(2-q_4,q_2+1,q-1,q+1;u,1-v) \bigg \} \,  .}
}
The results \Cone, with \Got, \thetwo, \thethree, \thefive, with \Ggt, and
\thesix\ give exactly the superconformal form with \OPER.

\medskip
\listrefs
\bye